\documentclass[sn-mathphys,Numbered]{sn-jnl}% Math and Physical Sciences Reference Style
\usepackage{graphicx}%
\usepackage{multirow}%
\usepackage{amsmath,amssymb,amsfonts}%
\usepackage{amsthm}%    
\usepackage{mathrsfs}%
\usepackage[title]{appendix}%
\usepackage{xcolor}%
\usepackage{textcomp}%
\usepackage{manyfoot}%
\usepackage{booktabs}%
\usepackage{algorithm}%
\usepackage{algorithmicx}%
\usepackage{algpseudocode}%
\usepackage{listings}%

\usepackage{tikz}

\usetikzlibrary{decorations.pathreplacing,calligraphy}

\newcommand{\MMDS}{{\sc Minimum Membership Dominating Set}}
\newcommand{\DS}{{\sc Dominating Set}}
\newcommand{\TDS}{{\sc Total Dominating Set}}
\newcommand{\SC}{{\sc Set Cover}}
\newcommand{\MSC}{{\sc Minimum Set Cover}}
\newcommand{\SCB}{{\sc \textbf{Set Cover}}}
\newcommand{\MMSC}{{\sc Minimum Membership Set Cover}}
\newcommand{\MMHS}{{\sc Minimum Membership Hitting Set}}
\newcommand{\MMDSB}{{\sc \textbf{Minimum Membership Dominating Set}}}

\begin{document}

\title[Algorithms for Minimum Membership Dominating Set Problem]{Algorithms for Minimum Membership Dominating Set Problem}


\author*[1]{\fnm{Sangam} \sur{Balchandar Reddy}}\email{21mcpc14@uohyd.ac.in}

\author[1]{\fnm{Anjeneya Swami} \sur{Kare}}\email{askcs@uohyd.ac.in}

\affil[1]{\orgdiv{School of Computer and Information Sciences}, \orgname{University of Hyderabad}, \orgaddress{\city{Hyderabad}, \postcode{500046}, \state{Telangana}, \country{India}}}


\abstract{Given a graph $G = (V, E)$ and an integer $k$, the \normalsize\MMDS{} \small(MMDS) problem asks to compute a set $S \subseteq V$ such that for each $v \in V$, $1 \leq |N[v] \cap S| \leq k$. The problem is known to be NP-complete even on split graphs and planar bipartite graphs. In this paper, we approach the problem from the algorithmic standpoint and obtain several interesting results. We give an $\mathcal{O}^*(1.747^n)$ time algorithm for the problem on split graphs. Following a reduction from a special case of 1-in-3 SAT problem, we show that there is no sub-exponential time algorithm running in time $\mathcal{O}^*(2^{o(n)})$ for bipartite graphs, for any $k \geq 2$. We also prove that the problem is NP-complete when $\Delta = k+2$, for any $k\geq 5$, even for bipartite graphs. We investigate the parameterized complexity of the problem for the parameter twin cover and the combined parameter distance to cluster, membership($k$) and prove that the problem is fixed-parameter tractable. Using a dynamic programming based approach, we obtain a linear-time algorithm for trees.}




\maketitle

\section{Introduction}\label{sec1}
Given a graph $G = (V, E)$, a set $D \subseteq V$ is a \textit{dominating set}, if each vertex $v \in V$ is either in $D$ or has a neighbour in $D$. Similarly, a set $D \subseteq V$ is a \textit{total dominating set}, if each vertex $v \in V$ has a neighbour in $D$. The \DS{} problem is one of Karp's 21 NP-complete problems \cite{Karp}. In this paper, we consider a variant of the \DS{} problem called \MMDS{} (MMDS). The MMDS problem seeks to compute a set $S \subseteq V$ such that for each vertex $v \in V$: $1 \leq |N[v] \cap S| \leq k.$ The problem is defined as follows.\vspace{3mm} \\
\textbf{\MMDSB{} (MMDS):} \vspace{1mm} \\
\textbf{Input:} A graph $G = (V, E)$ and a positive integer $k$. \\
\textbf{Parameter:} $k$ \\
\textbf{Question:} Does there exist a \textit{dominating set} $S \subseteq V$ such that $|N[v] \cap S| \leq k$ for each $v \in V$? \vspace{3mm} \\
MMDS is a decision problem and the term \textbf{Minimum} in \MMDS{} does not indicate that it is a minimization problem. We simply reuse the problem statement as defined in \cite{AA}. \vspace{3mm} \\
\textit{Known results.} Kuhn \cite{kuhn} initiated the study on the membership version of the \SC{} problem called \MMSC{}. They have proved that the problem is NP-complete and also obtained a lower bound on the approximation factor, which is $\mathcal{O}(\log n)$. Similarly, the \MMHS{} has been introduced by Narayanaswamy et al. \cite{NS}. In the geometric setting, they have shown that the problem does not admit a $2- \epsilon$ approximation algorithm for segments intersecting segments. They also provide a polynomial time algorithm for lines intersecting segments. Agrawal et al. \cite{AA} defined the MMDS problem and investigated its parameterized complexity. They have proved that the problem is NP-complete on planar bipartite graphs even for $k$=1. They have also shown that the problem admits an FPT algorithm for the parameter vertex cover. For the parameter pathwidth (hence for treewidth and clique-width), they have proved that the problem is W[1]-hard. \vspace{3mm} \\
The \DS{} problem is known to be NP-complete on split graphs and bipartite graphs \cite{Bertossi}. Fomin et al. \cite{Fomin2} gave exact algorithms for bipartite graphs and split graphs with a running time of $\mathcal{O}^*(1.73206^n)$ and $\mathcal{O}^*(1.41422^n)$ respectively. For general graphs, their approach leads to an exact algorithm with running time $\mathcal{O}^*(1.93782^n)$. Iwata \cite{Iwata} provided the best known exact algorithm for general graphs that runs in $\mathcal{O}^*(1.4864^n)$ time and polynomial space. The result was achieved by developing a new analyzing technique called the "potential method". Recently, Khzam \cite{Abu} presented an $\mathcal{O}^*(1.3384^n)$ algorithm for chordal graphs using the concept of simplicial vertices. Kikuno et al. \cite{Kikuno} proved that the \DS{} problem is NP-complete on cubic planar graphs. Alber et al. \cite{Alber}, obtained an $\mathcal{O}^*(4^{tw})$ time algorithm for \DS{} problem, where $tw$ is the treewidth of the graph. Later, Rooij et al. \cite{Rooij} proposed an algorithm with running time $\mathcal{O}^*(3^{tw})$. \vspace{3mm} \\
Given a graph $G = (V, E)$, $[i,j]$-\DS{} problem asks to compute a set $D \subseteq V$, such that each vertex in $V \setminus D$ has at least $i$ and at most $j$ neighbours in $D$. $[i,j]$-\TDS{} problem asks to compute a set $D \subseteq V$, such that each vertex in $V$ has at least $i$ and at most $j$ neighbours in $D$. Goharshady et al. \cite{Goharsh} proposed linear-time algorithms for $[1,2]$-\DS{} and $[1,2]$-\TDS{} problems on trees, which can be easily extended to $[i,j]$-variant of the problems. Chellali et al. \cite{CHELLALI} have proved that $[1, 2]$-Dominating set problem is NP-complete even on bipartite graphs. They have also provided bounds for various graph classes such as grid graphs, $P_4$-free graphs, and caterpillars. Meybodi et al. \cite{Meybodi} studied the parameterized complexity of $[1, j]$-\DS{} and $[1,j]$-\TDS{} problems and provided the lower bounds for split graphs (under ETH) and parameter pathwidth (under SETH). [1, $j$]-\DS{} and [1, $j$]-\TDS{} problems are known to be solvable in time $\mathcal{O}^*((j + 2)^{tw})$ and $\mathcal{O}^*((2j + 2)^{tw})$, respectively, on graphs width treewidth at most $tw$ \cite{Rooij}. \vspace{3mm} \\ 
\textit{Our results.} 
As the problem is NP-complete on planar bipartite graphs even for $k$=1, it is para-NP-hard for the membership parameter. Hence, we focus on various structural parameters and also provide exact algorithms. We obtain the following results.
\begin{enumerate}
    \item We provide an $\mathcal{O}^*(1.747^n)$ time exact algorithm for split graphs.
    \item Assuming ETH, we show that no sub-exponential time algorithm exists for the problem on bipartite graphs, for $k \geq 2$.
    \item We study the problem on bounded degree graphs and show that the problem is NP-complete when $\Delta = k+2$, for any $k\geq 5$, even for bipartite graphs.
    \item We investigate the parameterized complexity of the problem and prove that the problem is FPT when parameterized by twin cover and the combined parameter distance to cluster, membership($k$).
    \item Using a dynamic programming based approach, we provide a linear-time algorithm for trees.
\end{enumerate} \vspace{2mm}
\textit{Notations.} We consider only simple, finite, connected, and undirected graphs. Let $G = (V, E)$ be a graph with $V$ as the vertex set and $E$ as the edge set, such that $n = |V|$ and $m = |E|$. The set of vertices that belong to $N(v)$ and $N[v]$, respectively, are referred to as the neighbours and closed neighbours of a vertex $v$. The open neighbourhood of a set $T \subseteq V$ is denoted by $N(T)$ and the closed neighbourhood by $N[T]$. $N(T)$ = $\bigcup\limits_{v \in T}^{} N(v)$ and $N[T]$ = $\bigcup\limits_{v \in T}^{} N[v]$. The collective neighbourhood of a set $T$ in another set $D$ is the set of vertices in $N[T] \cap D$. Two vertices are said to be true twins, if they have the same closed neighbourhood. The degree of a vertex $v$ is represented by deg$(v)$ and deg$(v) = |N(v)|$. A graph is cubic if each vertex has a degree of three. Throughout this paper, we use $S$ to denote a \textit{minimum membership dominating set}. We say that a vertex $v \in V$ satisfies the MMDS constraint, if $1 \leq |N[v] \cap S| \leq k$. Similarly, we say that a set $T \subseteq V$ satisfies the MMDS constraint, if for every vertex $v \in T$, $1 \leq |N[v] \cap S| \leq k$. Other than this, we use the standard notations as defined in~\cite{WEST}. \vspace{3mm} \\
 We use $\mathcal{O}^*(f(n))$ to denote the time complexity of the form $\mathcal{O}(f(n) \cdot n^{\mathcal{O}(1)})$. A problem is considered to be \textit{fixed-parameter tractable} w.r.t. a parameter $k$, if there exists an algorithm with running time $\mathcal{O}^*(f(k))$, where $f$ is some computable function. Similarly, exact exponential algorithms run in time $\mathcal{O}^*(c^n)$ where $c$ is some constant and $c > 1$. For more information on \textit{parameterized complexity} and \textit{exact algorithms}, we refer the reader to~\cite{MC} and~\cite{fomin2013exact}, respectively. \vspace{3mm} \\
\section{Preliminary Results}\label{sec2}
In this section, assuming ETH, we show that a variant of the 3-SAT problem, that is, 3-CNF$^{\leq 3}$-XSAT does not admit an $\mathcal{O}^*(2^{o(n)})$ time algorithm. This result aids in proving some of the key results of the paper. \vspace{3mm} \\
We first define a SAT variant, XSAT as follows. \vspace{2mm} \\
\textbf{XSAT problem:} Given a CNF formula, the problem is to determine whether there exists an assignment such that each clause contains exactly one true literal. \vspace{3mm} \\
$k-$CNF: A CNF formula that contains at most $k$ literals in each clause.\vspace{3mm} \\
$k-$CNF$^{\leq l}:$ A CNF formula that contains at most $k$ literals in each clause and each variable occurs in at most $l$ clauses.\vspace{3mm} \\
$k-$CNF$_+$ and $k-$CNF$_+^{\leq l}$ have an additional constraint on $k$-CNF and $k$-CNF$^{\leq l}$, respectively, that all the literals are positive. \vspace{3mm} \\
We define the problems $k$-CNF-XSAT and $k$-CNF$^{\leq l}$-XSAT as follows: \vspace{2mm} \\
\textbf{$k$-CNF-XSAT problem:} Given a $k$-CNF formula, the problem is to determine whether there exists an assignment such that each clause contains exactly one true literal. \vspace{3mm} \\
\textbf{$k$-CNF$^{\leq l}$-XSAT problem:} Given a $k$-CNF formula with each variable occurring in at most $l$ clauses, the problem is to determine whether there exists an assignment such that each clause contains exactly one true literal. \vspace{3mm} \\
From \cite{Schafer}, we have that the problem $k-$CNF-XSAT is NP-complete for any value of $k\geq3$. There also exists a simpler transformation from 3-SAT to 3-CNF-XSAT. Consider an instance of the 3-SAT problem with $m$ clauses and $n$ variables. For a fixed value of $k$=3, we replace the clauses of the form: (1) ($x \vee y \vee z$) with ($\bar{x} \vee a \vee b$), ($y \vee b \vee c$) and ($\bar{z} \vee c \vee d$); (2) $(x \vee y)$ with ($\bar{x} \vee a \vee b$) and ($y \vee b$); clauses of 3-CNF-XSAT. Here, we use at most three clauses and at most four fresh variables for each clause of 3-SAT. We will have a total of at most 3$m$ clauses and at most $n+4m$ variables in the reduced instance of 3-CNF-XSAT. As the reduction preserves the linearity of size, we conclude that 3-CNF-XSAT also satisfies ETH. \vspace{3mm} \\ 
% Consider an instance of the 3-SAT problem with $m$ clauses and $n$ variables. Based on the reduction provided in Theorem 1, the reduced instance of XSAT for 3-CNF has size linear in the original instance. \vspace{3mm} \\
\textbf{Theorem 1.} \cite[Lemma 4.]{PORSCHEN} $k-$CNF$_+^{\leq l}$-XSAT and $k-$CNF$^{\leq l}$-XSAT remains NP-complete, for $k$,$l \geq$3. \qed \vspace{3mm} \\
Theorem 1 is proved by providing a polynomial time reduction from $k$-CNF-XSAT to $k$-CNF$^{\leq l}$-XSAT. Consider an instance of $k$-CNF-XSAT with $m$ clauses, $n$ variables and a total of at most $mk$ literals. According to Theorem 1, the reduced instance of $k$-CNF$^{\leq l}$-XSAT has at most $m+2mk$ clauses and $n+mk^2$ variables. For $k$=3, that becomes $7m$ clauses and $n+9m$ variables.\vspace{3mm} \\
\textbf{Theorem 2.} 3-CNF$^{\leq 3}$-XSAT problem can not be solved in $\mathcal{O}^*(2^{o(n)})$ time. \vspace{3mm} \\
\textit{Proof.} From Theorem 1, we have that 3-CNF$^{\leq 3}$-XSAT problem is NP-complete. There exists a reduction from 3-SAT to 3-CNF-XSAT with $3m$ clauses and $n+4m$ variables. The size of the reduced instance of the reduction from 3-CNF-XSAT to 3-CNF$^{\leq 3}$-XSAT is $7m$ clauses and $n+9m$ variables. The size of the reduced instance of 3-CNF$^{\leq 3}$-XSAT in terms of the size of $3$-SAT is $21m$ clauses and $n+31m$ variables. As the number of clauses and variables in the reduction are linear in the size of the original instance of 3-SAT; 3-CNF$^{\leq 3}$-XSAT also satisfies ETH. This concludes that, 3-CNF$^{\leq 3}$-XSAT problem can not be solved in $\mathcal{O}^*(2^{o(n)})$ time. \qed
\section{Exact algorithm for split graphs}
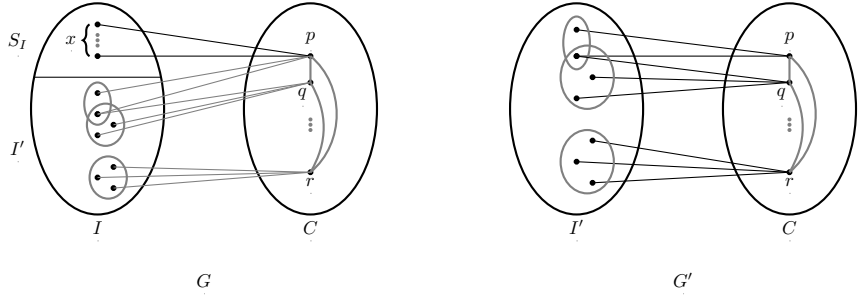
\begin{figure}
\centering
    \begin{tikzpicture} [thick,scale=0.7, every node/.style={scale=0.7}]
        \draw (0, 0) ellipse (1.25 and 2);
        \draw (4,0) ellipse (1.25 and 2);
        \draw[thin] (-1.2, 0.6) -- (1.2, 0.6);
        \filldraw (0, 1.6) circle (1pt) node[anchor=south]{};
        \filldraw (0, 1) circle (1pt) node[anchor=south]{};
        \filldraw[gray] (0, 1.4) circle (0.5pt) node[anchor=south]{};
        \filldraw[gray] (0, 1.2) circle (0.5pt) node[anchor=south]{};
        \filldraw[gray] (0, 1.3) circle (0.5pt) node[anchor=south]{};
        \filldraw (-1.5, 1) circle (0cm) node[anchor=south]{$S_I$};
        \filldraw (-1.5, -1) circle (0cm) node[anchor=south]{$I'$};
        \filldraw (0, -2.5) circle (0cm) node[anchor=south]{$I$};
        \filldraw (-0.5, 1.1) circle (0cm) node[anchor=south]{$x$};
        \filldraw (4, -2.5) circle (0cm) node[anchor=south]{$C$};
        \filldraw (2, -3.5) circle (0cm) node[anchor=south]{$G$};
        \draw[thin] (4, 1) -- (0, 1.6);
        \draw[thin] (4, 1) -- (0, 1);
        \draw [decorate, decoration = {brace}] (-0.15,1) --  (-0.15,1.6);

        \draw (9, 0) ellipse (1.25 and 2);
        \draw (13,0) ellipse (1.25 and 2);
        \filldraw (9, -2.5) circle (0cm) node[anchor=south]{$I'$};
        \filldraw (13, -2.5) circle (0cm) node[anchor=south]{$C$};
        \filldraw (11, -3.5) circle (0cm) node[anchor=south]{$G'$};
        \filldraw (13, 1.1) circle (0cm) node[anchor=south]{$p$};
        \filldraw (13, 1) circle (1pt) node[anchor=south]{};
        \filldraw (12.85, 0.05) circle (0cm) node[anchor=south]{$q$};
        \filldraw (13, 0.5) circle (1pt) node[anchor=south]{};
        \filldraw (13, -1.6) circle (0cm) node[anchor=south]{$r$};
        \filldraw (13, -1.2) circle (1pt) node[anchor=south]{};
        
        \filldraw (9, 1.5) circle (1pt);
        \filldraw (9, 1) circle (1pt); 
        
        \filldraw (9, 1) circle (1pt);
        \filldraw (9.3, 0.6) circle (1pt);      
        \filldraw (9, 0.2) circle (1pt);
        
        \filldraw (9.3, -1.4) circle (1pt);    
        \filldraw (9, -1) circle (1pt);
        \filldraw (9.3, -0.6) circle (1pt);

        \filldraw[gray] (13, -0.2) circle (0.5pt);    
        \filldraw[gray] (13, -0.3) circle (0.5pt);
        \filldraw[gray] (13, -0.4) circle (0.5pt);        

        \draw[thin] (9, 1.5) -- (13, 1);
        \draw[thin] (9, 1) -- (13, 1);
        \draw[gray] (9, 1.25) ellipse (0.25 and 0.5);
        
        \draw[thin] (9, 1) -- (13, 0.5);
        \draw[thin] (9.3, 0.6) -- (13, 0.5);        
        \draw[thin] (9, 0.2) -- (13, 0.5);
        \draw[gray] (9.2, 0.6) ellipse (0.5 and 0.6);
                
        \draw[thin] (9.3, -1.4) -- (13, -1.2);        
        \draw[thin] (9, -1) -- (13, -1.2);
        \draw[thin] (9.3, -0.6) -- (13, -1.2);
        \draw[gray] (9.2, -1) ellipse (0.5 and 0.6);

        \draw [gray] (13, 1) to (13, 0.5);
        \draw [gray] (13, 0.5) to [out=-60,in=60] (13, -1.2);
        \draw [gray] (13, 1) to [out=-40,in=40] (13, -1.2);

        \filldraw (4, 1.1) circle (0cm) node[anchor=south]{$p$};
        \filldraw (4, 1) circle (1pt) node[anchor=south]{};
        \filldraw (3.85, 0.05) circle (0cm) node[anchor=south]{$q$};
        \filldraw (4, 0.5) circle (1pt) node[anchor=south]{};
        \filldraw (4, -1.6) circle (0cm) node[anchor=south]{$r$};
        \filldraw (4, -1.2) circle (1pt) node[anchor=south]{};
        
        \draw [gray] (4, 1) to (4, 0.5);
        \draw [gray] (4, 0.5) to [out=-60,in=60] (4, -1.2);
        \draw [gray] (4, 1) to [out=-40,in=40] (4, -1.2);

        \filldraw[gray] (4, -0.2) circle (0.5pt);    
        \filldraw[gray] (4, -0.3) circle (0.5pt);
        \filldraw[gray] (4, -0.4) circle (0.5pt);   

        \filldraw (0, 0.3) circle (1pt);
        \filldraw (0, -0.1) circle (1pt); 
        
        \filldraw (0, -0.1) circle (1pt);
        \filldraw (0.3, -0.3) circle (1pt);      
        \filldraw (0, -0.5) circle (1pt);
        
        \filldraw (0.3, -1.1) circle (1pt);    
        \filldraw (0, -1.3) circle (1pt);
        \filldraw (0.3, -1.5) circle (1pt);   

        \draw[thin, gray] (0, 0.3) -- (4, 1);
        \draw[thin, gray] (0, -0.1) -- (4, 1);
        \draw[gray] (0, 0.1) ellipse (0.25 and 0.4);
        
        \draw[thin, gray] (0, -0.1) -- (4, 0.5);
        \draw[thin, gray] (0.3, -0.3) -- (4, 0.5);        
        \draw[thin, gray] (0, -0.5) -- (4, 0.5);
        \draw[gray] (0.15, -0.3) ellipse (0.35 and 0.4);
                
        \draw[thin, gray] (0.3, -1.1) -- (4, -1.2);        
        \draw[thin, gray] (0, -1.3) -- (4, -1.2);
        \draw[thin, gray] (0.3, -1.5) -- (4, -1.2);
        \draw[gray] (0.2, -1.3) ellipse (0.35 and 0.4);
        
    \end{tikzpicture}
    \caption{Posing a section of the MMDS problem as an instance of the \SC{}} 
    \label{fig:Fig1}
\end{figure}

A graph $G$ is a split graph if its vertices can be partitioned into $V1 \uplus V2$, such that $G[V_1]$ is a complete graph and $V_2$ is an independent set in $G$. In this section, we present an $\mathcal{O}^*(1.747^n)$ time algorithm for the MMDS problem on split graphs. We achieve this by posing a section of the problem as an instance of the \SC{} problem. \vspace{3mm} \\
\noindent \textbf{\SCB{} problem:}  \vspace{2mm} \\
Given a universe $\mathcal{U}$ and a family of sets $\mathcal{F}$, the \SC{} problem computes a minimum number of sets from $\mathcal{F}$ that covers $\mathcal{U}$. The decision version of the \SC{} problem is defined as follows. \vspace{3mm} \\
\textbf{Input:} A universe $\mathcal{U}$,  a family $\mathcal{F}$ over $\mathcal{U}$, and an integer $k$. \\
\textbf{Question:} Does there exists a subfamily $\mathcal{F'} \subseteq \mathcal{F}$ of size at most $k$ such that $\bigcup \mathcal{F'} = \mathcal{U}$?\vspace{3mm} \\
\textbf{Theorem 3.} \cite[Theorem 6.10.]{fomin2013exact} \textit{There exists an $\mathcal{O}(1.2353^{|\mathcal{U}| + |\mathcal{F}|})$ time exact algorithm to solve \MSC{} problem.} \qed \vspace{3mm} \\
Consider a split graph $G$. Let $I$ denote the independent set and $C$ denote the clique of $G$. If $|C| \leq k$, then the union of the vertices of $C$ and all the vertices of $I$ that have no neighbour in $C$ forms a \textit{minimum membership dominating set}. In this case, it will be an YES-instance. Hence, we assume that $|C| \geq k+1$. We define two sets $S_C$ and $S_I$, where $S_C$ = $S \cap C$ and $S_I$ = $S \cap I$. Even though the size of $C$ is more than $k$, $|S_C|$ will be at most $k$ (according to the MMDS constraint). We have the following two cases based on the size of $I$: (1) $|I| \leq \frac{n}{2}$ and (2) $|I| > \frac{n}{2}$. \vspace{2mm} \\ 
\textbf{Case 1:} $|I| \leq \frac{n}{2}$ and $|C| > \frac{n}{2}$. \vspace{2mm} \\
We guess the vertex set $S_I$ of size ranging from 0 to $\frac{n}{2}$. This can be done in $\mathcal{O}(2^{\frac{n}{2}})$ ways. We find a vertex from $C$ that has maximum number of neighbours in $S_I$. Let this vertex be $p$ and let the number of neighbours of it in $S_I$ be $x$. $p$ is adjacent to $x$ vertices from $S_I$ and in order to satisfy the MMDS constraint for $p$, $|S_C|$ must be at most $k-x$. We remove the vertex set that is guessed to be $S_I$. We have a new instance $G'$ in which the independent set $I'$ has only undominated vertices and some subset of vertices of $C$ are dominated. In the end, we only have at most $k-x$ vertices in $S_C$, and no vertices from $G'$ will violate the MMDS constraint. \vspace{2mm} \\
We pose this problem as a \SC{} instance as follows. See \autoref{fig:Fig1} for an illustration. \vspace{2mm} \\
\textbf{Input:} A universe $\mathcal{U} = I'$ and sets $\mathcal{F} = \{N(v)\cap I': v \in C\}$. \\
\textbf{Question:} Is there an $\mathcal{F'} \subseteq \mathcal{F}$ of size at most $k-x$, such that $\bigcup \mathcal{F'} = \mathcal{U}$? \vspace{2mm} \\
\setlength{\textfloatsep}{1\baselineskip plus 0\baselineskip minus 0\baselineskip}
Each set in the \SC{} instance corresponds to a unique vertex in $C$ and the size of the universe is $I'$, which is at most the size of $I$. Therefore, we have $|\mathcal{F}| + |\mathcal{U}| \leq n$. From Theorem 3, we check whether there exists a \textit{set cover} of size $k-x$ in $\mathcal{O}^*(1.2353^n)$ time. The vertices of $C$ that correspond to the sets $\mathcal{F'}$ will form $S_C$. \vspace{2mm} \\
\textbf{Case 2:} $|I| > \frac{n}{2}$ and $k+1 \leq |C| \leq \frac{n}{2}$. \vspace{2mm} \\ We guess the vertex set $S_C$ of size ranging from 0 to $k$. This can be done in $\mathcal{O}(2^{\frac{n}{2}})$ ways. Once $S_C$ is fixed, we compute $S_I$. Let $|S_C| \geq 1$. At this point, all the vertices of $C$ are dominated. The vertices from $I$ with no neighbour in $S_C$ are not yet dominated. All these vertices must dominate themselves. Hence, they form the set $S_I$. Let $|S_C|=0$. As all the vertices of $I$ are undominated, they must be a part of $S$. Therefore, $S_I$ contains all the vertices of $I$. \vspace{2mm} \\ 
Once $S_I$ and $S_C$ are fixed, we check whether $S_I \cup S_C$ forms a \textit{minimum membership dominating set}, in polynomial time. Case 1 takes $\mathcal{O}^*(2^\frac{n}{2} \cdot 1.2353^n)$ time and Case 2 takes $\mathcal{O}^*(2^\frac{n}{2})$ time. The overall time complexity is $\mathcal{O}^*(2^\frac{n}{2} \cdot 1.2353^n)$ time, that is $\mathcal{O}^*(1.747^n)$. \vspace{3mm} \\ 
To summarize, we have the following theorem. \vspace{2mm} \\
\textbf{Theorem 4. } \textit{The }\MMDS{}\textit{ problem on split graphs with $n$ vertices can be solved in $\mathcal{O}^*(1.747^n)$ time.} \qed

\section{Complexity on bipartite graphs}
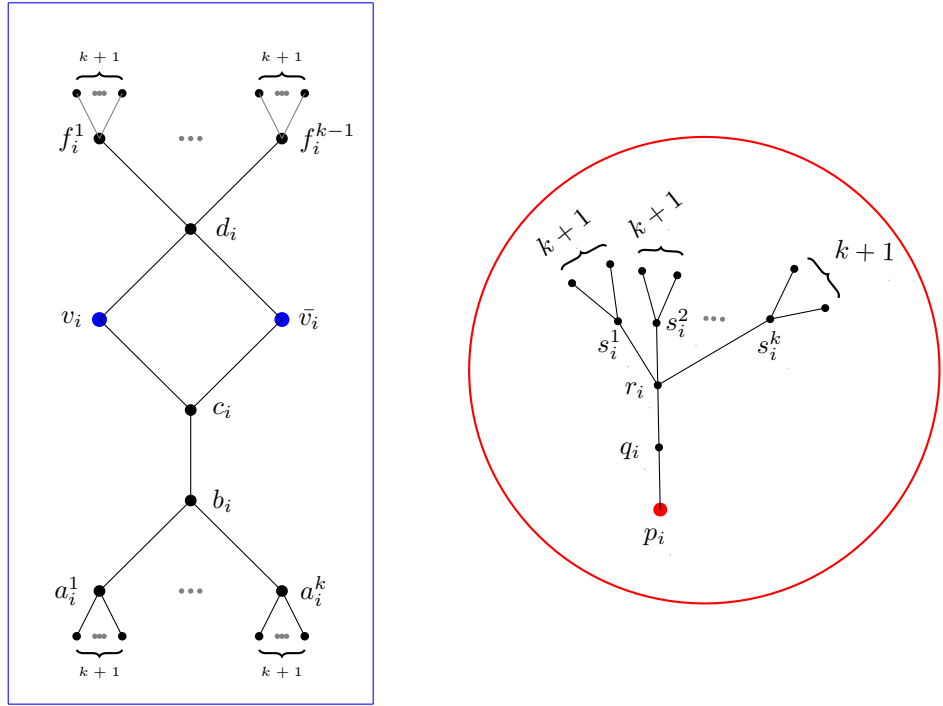
\begin{figure}
    \centering
    \begin{minipage}[b]{0.05\textwidth}
    \begin{tikzpicture} 
    \end{tikzpicture}
    \end{minipage}
    \hfill
    \begin{minipage}[b]{0.25\textwidth}
    \begin{tikzpicture} [thick,scale=1.2, every node/.style={scale=1}]
    \filldraw [blue] (5, 9) circle (2pt);
    \filldraw [blue] (7, 9) circle (2pt);
    % \filldraw [black] (4.5, 10) circle (1.5pt);
    % \filldraw [gray] (4.9, 10) circle (0.5pt);
    % \filldraw [gray] (5, 10) circle (0.5pt);
    % \filldraw [gray] (5.1, 10) circle (0.5pt);
    % \filldraw [black] (5.5, 10) circle (1.5pt);
    \filldraw [black] (6, 10) circle (1.5pt);
    \filldraw [black] (5, 11) circle (1.5pt);
    \filldraw [gray] (5.9, 11) circle (0.5pt);
    \filldraw [gray] (6, 11) circle (0.5pt);
    \filldraw [gray] (6.1, 11) circle (0.5pt);
    \filldraw [black] (7, 11) circle (1.5pt);
    
    \filldraw [black] (6, 8) circle (1.5pt);

    \draw[black, thin] (6, 8) -- (6, 7);  
    
    \draw[black, thin] (5, 9) -- (6, 10);
    \draw[black, thin] (7, 9) -- (6, 10);
    % \draw[black, thin] (5, 9) -- (5, 8);
    \draw[black, thin] (5, 9) -- (6, 8);
    % \draw[black, thin] (7, 9) -- (5, 8);
    \draw[black, thin] (7, 9) -- (6, 8);
    \draw[black, thin] (6, 10) -- (5, 11);
    \draw[black, thin] (6, 10) -- (7, 11);

    % \filldraw [black] (4.25, 10.5) circle (1pt);
    % \filldraw [gray] (4.45, 10.5) circle (0.5pt);
    % \filldraw [gray] (4.5, 10.5) circle (0.5pt);
    % \filldraw [gray] (4.55, 10.5) circle (0.5pt);
    % \filldraw [black] (4.75, 10.5) circle (1pt);
    
    % \filldraw [black] (5.25, 10.5) circle (1pt);
    % \filldraw [gray] (5.45, 10.5) circle (0.5pt);
    % \filldraw [gray] (5.5, 10.5) circle (0.5pt);
    % \filldraw [gray] (5.55, 10.5) circle (0.5pt);
    % \filldraw [black] (5.75, 10.5) circle (1pt);
    
    \filldraw [black] (4.75, 11.5) circle (1pt);
    \filldraw [gray] (4.95, 11.5) circle (0.5pt);
    \filldraw [gray] (5, 11.5) circle (0.5pt);
    \filldraw [gray] (5.05, 11.5) circle (0.5pt);
    \filldraw [black] (5.25, 11.5) circle (1pt);

    \filldraw [black] (6.75, 11.5) circle (1pt);
    \filldraw [gray] (6.95, 11.5) circle (0.5pt);
    \filldraw [gray] (7, 11.5) circle (0.5pt);
    \filldraw [gray] (7.05, 11.5) circle (0.5pt);
    \filldraw [black] (7.25, 11.5) circle (1pt);
    
    \draw[gray, thin] (5, 11) -- (5.25, 11.5);
    \draw[gray, thin] (5, 11) -- (4.75, 11.5);
    \draw[gray, thin] (7, 11) -- (7.25, 11.5);
    \draw[gray, thin] (7, 11) -- (6.75, 11.5); 
    
    % \node[] at (4.3, 9.8) {$a_i^1$};
    % \node[] at (6.1, 10) {$a_i^{k-1}$};
    \node[] at (6.4, 10) {$d_i$};
    \node[] at (4.7, 9) {$v_i$};
    \node[] at (7.3, 9) {$\bar{v_i}$};
    % \node[] at (4.7, 8) {$b_i^1$};
    \node[] at (6.35, 8) {$c_i$};    
    \node[] at (4.7, 11) {$f_i^1$};
    \node[] at (7.5, 11) {$f_i^{k-1}$};

    \draw [decorate, decoration = {brace}] (4.75, 11.65) --  (5.25,11.65);
    \draw [decorate, decoration = {brace}] (6.75,11.65) --  (7.25,11.65);

    \node[] at (5, 11.9) {\tiny{$k+1$}};
    \node[] at (7, 11.9) {\tiny{$k+1$}};

    \node[] at (5, 5.1) {\tiny{$k+1$}};
    \node[] at (7, 5.1) {\tiny{$k+1$}};
    
    \filldraw [black] (6, 7) circle (1.5pt);

    \filldraw [black] (5, 6) circle (1.5pt);
    \filldraw [gray] (5.9, 6) circle (0.5pt);
    \filldraw [gray] (6, 6) circle (0.5pt);
    \filldraw [gray] (6.1, 6) circle (0.5pt);
    \filldraw [black] (7, 6) circle (1.5pt);

    \filldraw [black] (4.75, 5.5) circle (1pt);
    \filldraw [gray] (4.95, 5.5) circle (0.5pt);
    \filldraw [gray] (5, 5.5) circle (0.5pt);
    \filldraw [gray] (5.05, 5.5) circle (0.5pt);
    \filldraw [black] (5.25, 5.5) circle (1pt);

    \filldraw [black] (6.75, 5.5) circle (1pt);
    \filldraw [gray] (6.95, 5.5) circle (0.5pt);
    \filldraw [gray] (7, 5.5) circle (0.5pt);
    \filldraw [gray] (7.05, 5.5) circle (0.5pt);
    \filldraw [black] (7.25, 5.5) circle (1pt);

    \node[] at (6.35, 7) {$b_i$}; 
    \node[] at (4.65, 6) {$a_i^1$};  
    \node[] at (7.35, 6) {$a_i^k$};

    \draw [decorate, decoration = {brace}] (5.25, 5.35) --  (4.75, 5.35);
    \draw [decorate, decoration = {brace}] (7.25, 5.35) --  (6.75, 5.35);

    \draw[black, thin] (6, 7) -- (5, 6);    
    \draw[black, thin] (6, 7) -- (7, 6);
    
    \draw[black, thin] (5, 6) -- (4.75, 5.5);    
    \draw[black, thin] (5, 6) -- (5.25, 5.5);
    
    \draw[black, thin] (7, 6) -- (6.75, 5.5); 
    \draw[black, thin] (7, 6) -- (7.25, 5.5);  

    \draw[blue, thin] (4, 4.75) -- (8, 4.75);    
    \draw[blue, thin] (8, 4.75) -- (8, 12.5);
    \draw[blue, thin] (4, 4.75) -- (4, 12.5);    
    \draw[blue, thin] (8, 12.5) -- (4, 12.5);
    
    \end{tikzpicture}
    \end{minipage}    
    \hfill
    \begin{minipage}[b]{0.05\textwidth}
    \begin{tikzpicture} 
    \end{tikzpicture}
    \end{minipage}
    \hfill
    \begin{minipage}[b]{0.4\textwidth}
    \rotatebox{38}
    {
    \begin{tikzpicture} [thick,scale=1.1, every node/.style={scale=1}]
   \filldraw [black] (13.15, 12.7) circle (1pt);
    \filldraw (12.9, 12.55) circle (0cm) node[anchor=south]{\rotatebox{322}{$r_i$}};
    \filldraw [black] (13.6, 13.3) circle (1pt);
    \filldraw (13.8, 12.8) circle (0cm) node[anchor=south]{\rotatebox{322}{$s_i^2$}};
    \filldraw [black] (13.25, 13.6) circle (1pt);
    \filldraw (13, 13.1) circle (0cm) node[anchor=south]{\rotatebox{322}{$s_i^1$}};
    \filldraw [black] (14.7, 12.5) circle (1pt);
    \filldraw (14.5, 11.9) circle (0cm) node[anchor=south]{\rotatebox{322}{$s_i^k$}};

    \draw[black, thin] (12.7, 12.1) -- (13.15, 12.7);
    \draw[black, thin] (13.15, 12.7) -- (13.6, 13.3);    
    \draw[black, thin] (13.15, 12.7) -- (13.25, 13.6);    
    \draw[black, thin] (13.15, 12.7) -- (14.7, 12.5);
    
    \filldraw [black] (13.1, 14.3) circle (1pt);
    \filldraw [black] (13.6, 14.2) circle (1pt);
    \filldraw [black] (14.15, 13.6) circle (1pt);
    \filldraw [black] (13.85, 13.9) circle (1pt);
    \filldraw [black] (15.3, 12.2) circle (1pt);
    \filldraw [black] (15.3, 12.8) circle (1pt);

    \filldraw [gray] (14.1, 12.98) circle (0.5pt);
    \filldraw [gray] (14.17, 12.92) circle (0.5pt);
    \filldraw [gray] (14.25, 12.85) circle (0.5pt);

    \draw[black, thin] (13.25, 13.6) -- (13.1, 14.3);
    \draw[black, thin] (13.25, 13.6) -- (13.6, 14.2);  
    
    \draw [decorate, decoration = {brace}] (13.1, 14.5) --  (13.7, 14.4);
    \filldraw (13.4, 14.6) circle (0cm) node[anchor=south]{$k+1$};
    
    \draw[black, thin] (13.6, 13.3) -- (14.15, 13.6);   
    \draw[black, thin] (13.6, 13.3) -- (13.85, 13.9);
    
    \draw [decorate, decoration = {brace}]  (13.95, 14.1) -- (14.35, 13.75);
    \filldraw (14.4, 14.1) circle (0cm) node[anchor=south]{$k+1$};
    
    \draw[black, thin] (14.7, 12.5) -- (15.3, 12.2);   
    \draw[black, thin] (14.7, 12.5) -- (15.3, 12.8);
    
    \draw [decorate, decoration = {brace}] (15.5, 12.8) --  (15.5, 12.2);
    \filldraw (16.1, 12) circle (0cm) node[anchor=south]{\rotatebox{320}{$k+1$}};

    \filldraw [red] (12.25, 11.5) circle (2pt);
    \filldraw (12, 11) circle (0cm) node[anchor=south]{\rotatebox{322}{$p_i$}};
    \filldraw [black] (12.7, 12.1) circle (1pt);
    \filldraw (12.4, 12) circle (0cm) node[anchor=south]{\rotatebox{322}{$q_i$}};
    \draw[black, thin] (12.7, 12.1) -- (12.25, 11.5);
    \draw [red] (13.7, 12.5) circle (80pt);
    \end{tikzpicture}}
    
    \end{minipage} 
    \hfill   
    \begin{minipage}[b]{0.1\textwidth}
    \begin{tikzpicture} 
    \end{tikzpicture}
    \end{minipage}
    \hfill \vspace{4mm} \\
    \caption{ (a) Variable gadget for some $i \in [n]$ (on the left). (b) Literal gadget for some $i \in [n]$ (on the right).}
    \label{fig: Fig4}
\end{figure}

\begin{figure} [t]
    \centering
    \begin{tikzpicture} [thick,scale=0.85, every node/.style={scale=0.85}]
    \filldraw [black] (5, 9) circle (2pt);
    \filldraw [black] (6, 9) circle (2pt);

    \filldraw [black] (4, 9) circle (2pt);
    \filldraw [black] (7, 9) circle (2pt);
    
    \draw[gray] (4, 9) -- (5, 9);
    \draw[gray] (7, 9) -- (6, 9);

    \node[] at (3.85, 8.75) {$p_1$};
    \node[] at (7.25, 8.75) {$\bar{p_1}$};
    
    \filldraw [black] (7, 6) circle (1.5pt);
    \filldraw [black] (10.4, 6) circle (1.5pt);
    \filldraw [black] (14, 6) circle (1.5pt);

    \filldraw [black] (7, 12) circle (1.5pt);
    \filldraw [black] (10.4, 12) circle (1.5pt);
    \filldraw [black] (14, 12) circle (1.5pt);

    \filldraw [black] (10, 9) circle (2pt);
    \filldraw [black] (11, 9) circle (2pt);

    \filldraw [black] (9, 9) circle (2pt);
    \filldraw [black] (12, 9) circle (2pt);
    
    \draw[gray] (10, 9) -- (9, 9);
    \draw[gray] (11, 9) -- (12, 9);

    \node[] at (8.85, 8.75) {$p_2$};
    \node[] at (12.25, 8.75) {$\bar{p_2}$};
    
    \filldraw [black] (14.5, 9) circle (2pt);
    \filldraw [black] (15.5, 9) circle (2pt);

    \filldraw [black] (13.5, 9) circle (2pt);
    \filldraw [black] (16.5, 9) circle (2pt);
    
    \draw[gray] (13.5, 9) -- (14.5, 9);
    \draw[gray] (15.5, 9) -- (16.5, 9);

    \node[] at (13.35, 8.75) {$p_3$};
    \node[] at (16.75, 8.75) {$\bar{p_3}$};

    \draw[gray] (7, 6) -- (5, 9);
    \draw[gray] (10.4, 6) -- (6, 9);
    \draw[gray] (7, 6) -- (11, 9);
    \draw[gray] (14, 6) -- (10, 9);
    \draw[gray] (10.4, 6) -- (14.5, 9);
    \draw[gray] (14, 6) -- (15.5, 9);
    \draw[gray] (7, 6) -- (14.5, 9);
    \draw[gray] (10.4, 6) -- (11, 9);
    \draw[gray] (14, 6) -- (6, 9);

    \draw[gray] (7, 12) -- (5, 9);
    \draw[gray] (10.4, 12) -- (6, 9);
    \draw[gray] (7, 12) -- (11, 9);
    \draw[gray] (14, 12) -- (10, 9);
    \draw[gray] (10.4, 12) -- (14.5, 9);
    \draw[gray] (14, 12) -- (15.5, 9);
    \draw[gray] (7, 12) -- (14.5, 9);
    \draw[gray] (10.4, 12) -- (11, 9);
    \draw[gray] (14, 12) -- (6, 9);

    \node[] at (4.85, 8.75) {$v_1$};
    \node[] at (5.85, 8.75) {$\bar{v_1}$};

    \node[] at (9.85, 8.75) {$v_2$};
    \node[] at (11.25, 8.75) {$\bar{v_2}$};
    \node[] at (14.65, 8.75) {$v_3$};
    \node[] at (15.75, 8.75) {$\bar{v_3}$};

    \node[] at (6.7, 6) {$u_1$};
    \node[] at (10.8, 6) {$u_2$};
    \node[] at (14.5, 6) {$u_3$};

    \node[] at (6.7, 12) {$w_1$};
    \node[] at (10.8, 12) {$w_2$};
    \node[] at (14.5, 12) {$w_3$};

    \node[] at (5.7, 12.95) {$h_1^1$};
    \node[] at (8.5, 12.85) {$h_1^{k-1}$};

    \node[] at (9.2, 12.8) {$h_2^1$};
    \node[] at (11.9, 12.8) {$h_2^{k-1}$};

    \node[] at (12.75, 12.8) {$h_3^1$};
    \node[] at (15.5, 12.95) {$h_3^{k-1}$};

    \node[] at (16, 5) {};

    \filldraw [black] (6, 13) circle (1.5pt);
    \filldraw [gray] (6.9, 13) circle (0.5pt);
    \filldraw [gray] (7, 13) circle (0.5pt);
    \filldraw [gray] (7.1, 13) circle (0.5pt);
    \filldraw [black] (8, 13) circle (1.5pt);

    \draw[thin, black] (7, 12) -- (6, 13);
    \draw[thin, black] (7, 12) -- (8, 13);

    \filldraw [black] (5.75, 13.5) circle (1pt);
    \filldraw [gray] (5.95, 13.5) circle (0.5pt);
    \filldraw [gray] (6, 13.5) circle (0.5pt);
    \filldraw [gray] (6.05, 13.5) circle (0.5pt);
    \filldraw [black] (6.25, 13.5) circle (1pt);

    \draw[thin, black] (6, 13) -- (5.75, 13.5);
    \draw[thin, black] (6, 13) -- (6.25, 13.5);

    \filldraw [black] (7.75, 13.5) circle (1pt);
    \filldraw [gray] (7.95, 13.5) circle (0.5pt);
    \filldraw [gray] (8, 13.5) circle (0.5pt);
    \filldraw [gray] (8.05, 13.5) circle (0.5pt);
    \filldraw [black] (8.25, 13.5) circle (1pt);

    \draw[thin, black] (8, 13) -- (7.75, 13.5);
    \draw[thin, black] (8, 13) -- (8.25, 13.5);

    \filldraw [black] (9.4, 13) circle (1.5pt);
    \filldraw [gray] (10.3, 13) circle (0.5pt);
    \filldraw [gray] (10.4, 13) circle (0.5pt);
    \filldraw [gray] (10.5, 13) circle (0.5pt);
    \filldraw [black] (11.4, 13) circle (1.5pt);

    \draw[thin, black] (10.4, 12) -- (9.4, 13);
    \draw[thin, black] (10.4, 12) -- (11.4, 13);

    \filldraw [black] (9.15, 13.5) circle (1pt);
    \filldraw [gray] (9.35, 13.5) circle (0.5pt);
    \filldraw [gray] (9.4, 13.5) circle (0.5pt);
    \filldraw [gray] (9.45, 13.5) circle (0.5pt);
    \filldraw [black] (9.65, 13.5) circle (1pt);

    \draw[thin, black] (9.4, 13) -- (9.15, 13.5);
    \draw[thin, black] (9.4, 13) -- (9.65, 13.5);

    \filldraw [black] (11.15, 13.5) circle (1pt);
    \filldraw [gray] (11.35, 13.5) circle (0.5pt);
    \filldraw [gray] (11.4, 13.5) circle (0.5pt);
    \filldraw [gray] (11.45, 13.5) circle (0.5pt);
    \filldraw [black] (11.65, 13.5) circle (1pt);

    \draw[thin, black] (11.4, 13) -- (11.15, 13.5);
    \draw[thin, black] (11.4, 13) -- (11.65, 13.5);

    \filldraw [black] (13, 13) circle (1.5pt);
    \filldraw [gray] (13.9, 13) circle (0.5pt);
    \filldraw [gray] (14, 13) circle (0.5pt);
    \filldraw [gray] (14.1, 13) circle (0.5pt);
    \filldraw [black] (15, 13) circle (1.5pt);

    \draw[thin, black] (14, 12) -- (13, 13);
    \draw[thin, black] (14, 12) -- (15, 13);

    \filldraw [black] (12.75, 13.5) circle (1pt);
    \filldraw [gray] (12.95, 13.5) circle (0.5pt);
    \filldraw [gray] (13, 13.5) circle (0.5pt);
    \filldraw [gray] (13.05, 13.5) circle (0.5pt);
    \filldraw [black] (13.25, 13.5) circle (1pt);

    \draw[thin, black] (13, 13) -- (12.75, 13.5);
    \draw[thin, black] (13, 13) -- (13.25, 13.5);

    \filldraw [black] (14.75, 13.5) circle (1pt);
    \filldraw [gray] (14.95, 13.5) circle (0.5pt);
    \filldraw [gray] (15, 13.5) circle (0.5pt);
    \filldraw [gray] (15.05, 13.5) circle (0.5pt);
    \filldraw [black] (15.25, 13.5) circle (1pt);

    \draw[thin, black] (15, 13) -- (14.75, 13.5);
    \draw[thin, black] (15, 13) -- (15.25, 13.5);

    \draw [decorate, decoration = {brace}] (5.7,13.65) --  (6.3,13.65);
    \draw [decorate, decoration = {brace}] (7.7,13.65) --  (8.3,13.65);

    \draw [decorate, decoration = {brace}] (9.1,13.65) --  (9.7,13.65);
    \draw [decorate, decoration = {brace}] (11.1,13.65) --  (11.7,13.65);

    \draw [decorate, decoration = {brace}] (12.7,13.65) --  (13.3,13.65);
    \draw [decorate, decoration = {brace}] (14.7,13.65) --  (15.3,13.65);

    \node[] at (6, 14) {\tiny{$k+1$}};
    \node[] at (8, 14) {\tiny{$k+1$}};

    \node[] at (9.5, 14) {\tiny{$k+1$}};
    \node[] at (11.5, 14) {\tiny{$k+1$}};

    \node[] at (13, 14) {\tiny{$k+1$}};
    \node[] at (15, 14) {\tiny{$k+1$}};

    %iteration 1
    
    \node[] at (6, 3.1) {\tiny{$k+1$}};
    \node[] at (8, 3.1) {\tiny{$k+1$}};
    
    \filldraw [black] (7, 5) circle (1.5pt);

    \filldraw [black] (6, 4) circle (1.5pt);
    \filldraw [gray] (6.9, 4) circle (0.5pt);
    \filldraw [gray] (7, 4) circle (0.5pt);
    \filldraw [gray] (7.1, 4) circle (0.5pt);
    \filldraw [black] (8, 4) circle (1.5pt);

    \filldraw [black] (5.75, 3.5) circle (1pt);
    \filldraw [gray] (5.95, 3.5) circle (0.5pt);
    \filldraw [gray] (6, 3.5) circle (0.5pt);
    \filldraw [gray] (6.05, 3.5) circle (0.5pt);
    \filldraw [black] (6.25, 3.5) circle (1pt);

    \filldraw [black] (7.75, 3.5) circle (1pt);
    \filldraw [gray] (7.95, 3.5) circle (0.5pt);
    \filldraw [gray] (8, 3.5) circle (0.5pt);
    \filldraw [gray] (8.05, 3.5) circle (0.5pt);
    \filldraw [black] (8.25, 3.5) circle (1pt);

    \node[] at (7.35, 5) {$y_1$}; 
    \node[] at (5.65, 4) {$z_1^1$};  
    \node[] at (8.35, 4) {$z_1^k$};

    \draw [decorate, decoration = {brace}] (6.25, 3.35) --  (5.75, 3.35);
    \draw [decorate, decoration = {brace}] (8.25, 3.35) --  (7.75, 3.35);

    \draw[black, thin] (7, 6) -- (7, 5);
    \draw[black, thin] (7, 5) -- (6, 4);    
    \draw[black, thin] (7, 5) -- (8, 4);
    
    \draw[black, thin] (6, 4) -- (5.75, 3.5);    
    \draw[black, thin] (6, 4) -- (6.25, 3.5);
    
    \draw[black, thin] (8, 4) -- (7.75, 3.5); 
    \draw[black, thin] (8, 4) -- (8.25, 3.5); 
    
    %iteration 2

    \node[] at (9.4, 3.1) {\tiny{$k+1$}};
    \node[] at (11.4, 3.1) {\tiny{$k+1$}};
    
    \filldraw [black] (10.4, 5) circle (1.5pt);

    \filldraw [black] (9.4, 4) circle (1.5pt);
    \filldraw [gray] (10.3, 4) circle (0.5pt);
    \filldraw [gray] (10.4, 4) circle (0.5pt);
    \filldraw [gray] (10.5, 4) circle (0.5pt);
    \filldraw [black] (11.4, 4) circle (1.5pt);

    \filldraw [black] (9.15, 3.5) circle (1pt);
    \filldraw [gray] (9.35, 3.5) circle (0.5pt);
    \filldraw [gray] (9.4, 3.5) circle (0.5pt);
    \filldraw [gray] (9.45, 3.5) circle (0.5pt);
    \filldraw [black] (9.65, 3.5) circle (1pt);

    \filldraw [black] (11.15, 3.5) circle (1pt);
    \filldraw [gray] (11.35, 3.5) circle (0.5pt);
    \filldraw [gray] (11.4, 3.5) circle (0.5pt);
    \filldraw [gray] (11.45, 3.5) circle (0.5pt);
    \filldraw [black] (11.65, 3.5) circle (1pt);

    \node[] at (10.75, 5) {$y_2$}; 
    \node[] at (9.05, 4) {$z_2^1$};  
    \node[] at (11.75, 4) {$z_2^k$};

    \draw [decorate, decoration = {brace}] (9.65, 3.35) --  (9.15, 3.35);
    \draw [decorate, decoration = {brace}] (11.65, 3.35) --  (11.15, 3.35);

    \draw[black, thin] (10.4, 6) -- (10.4, 5);
    \draw[black, thin] (10.4, 5) -- (9.4, 4);    
    \draw[black, thin] (10.4, 5) -- (11.4, 4);
    
    \draw[black, thin] (9.4, 4) -- (9.15, 3.5);    
    \draw[black, thin] (9.4, 4) -- (9.65, 3.5);
    
    \draw[black, thin] (11.4, 4) -- (11.15, 3.5); 
    \draw[black, thin] (11.4, 4) -- (11.65, 3.5); 
    %iteration 3
    
    \node[] at (13, 3.1) {\tiny{$k+1$}};
    \node[] at (15, 3.1) {\tiny{$k+1$}};
    
    \filldraw [black] (14, 5) circle (1.5pt);

    \filldraw [black] (13, 4) circle (1.5pt);
    \filldraw [gray] (13.9, 4) circle (0.5pt);
    \filldraw [gray] (14, 4) circle (0.5pt);
    \filldraw [gray] (14.1, 4) circle (0.5pt);
    \filldraw [black] (15, 4) circle (1.5pt);

    \filldraw [black] (12.75, 3.5) circle (1pt);
    \filldraw [gray] (12.95, 3.5) circle (0.5pt);
    \filldraw [gray] (13, 3.5) circle (0.5pt);
    \filldraw [gray] (13.05, 3.5) circle (0.5pt);
    \filldraw [black] (13.25, 3.5) circle (1pt);

    \filldraw [black] (14.75, 3.5) circle (1pt);
    \filldraw [gray] (14.95, 3.5) circle (0.5pt);
    \filldraw [gray] (15, 3.5) circle (0.5pt);
    \filldraw [gray] (15.05, 3.5) circle (0.5pt);
    \filldraw [black] (15.25, 3.5) circle (1pt);

    \node[] at (14.35, 5) {$y_3$}; 
    \node[] at (12.65, 4) {$z_3^1$};  
    \node[] at (15.35, 4) {$z_3^k$};

    \draw [decorate, decoration = {brace}] (13.25, 3.35) --  (12.75, 3.35);
    \draw [decorate, decoration = {brace}] (15.25, 3.35) --  (14.75, 3.35);

    \draw[black, thin] (14, 6) -- (14, 5);
    \draw[black, thin] (14, 5) -- (13, 4);    
    \draw[black, thin] (14, 5) -- (15, 4);
    
    \draw[black, thin] (13, 4) -- (12.75, 3.5);    
    \draw[black, thin] (13, 4) -- (13.25, 3.5);
    
    \draw[black, thin] (15, 4) -- (14.75, 3.5); 
    \draw[black, thin] (15, 4) -- (15.25, 3.5); 

    \draw[blue, thin] (4.6, 9.5) -- (6.4, 9.5); 
    \draw[blue, thin] (4.6, 8.5) -- (6.4, 8.5); 
    \draw[blue, thin] (4.6, 9.5) -- (4.6, 8.5); 
    \draw[blue, thin] (6.4, 8.5) -- (6.4, 9.5); 

    \draw[blue, thin] (9.6, 9.5) -- (11.5, 9.5); 
    \draw[blue, thin] (9.6, 8.5) -- (11.5, 8.5); 
    \draw[blue, thin] (9.6, 9.5) -- (9.6, 8.5); 
    \draw[blue, thin] (11.5, 8.5) -- (11.5, 9.5); 

    \draw[blue, thin] (14.1, 9.5) -- (16, 9.5); 
    \draw[blue, thin] (14.1, 8.5) -- (16, 8.5); 
    \draw[blue, thin] (14.1, 9.5) -- (14.1, 8.5); 
    \draw[blue, thin] (16, 8.5) -- (16, 9.5);

    \draw [red] (3.7, 9) circle (15pt);
    \draw [red] (7.2, 9) circle (15pt);
    
    \draw [red] (8.7, 9) circle (15pt);
    \draw [red] (12.1, 9) circle (15pt);
    
    \draw [red] (13.3, 9) circle (15pt);
    \draw [red] (16.8, 9) circle (15pt);
    
    \end{tikzpicture}
    \caption{Construction of an MMDS instance from the 3-CNF$^{\leq 3}$-XSAT formula: 
$\phi = (x_1 \vee \bar{x_2} \vee x_3) \wedge (\bar{x_1} \vee \bar{x_2} \vee x_3) \wedge  (\bar{x_1} \vee x_2 \vee \bar{x_3})$. Here, the red circles and blue rectangles are placeholders for literal gadget and variable gadgets. The literal gadget and variable gadget are illustrated in \autoref{fig: Fig4} }
    \label{fig: Fig7}
\end{figure}
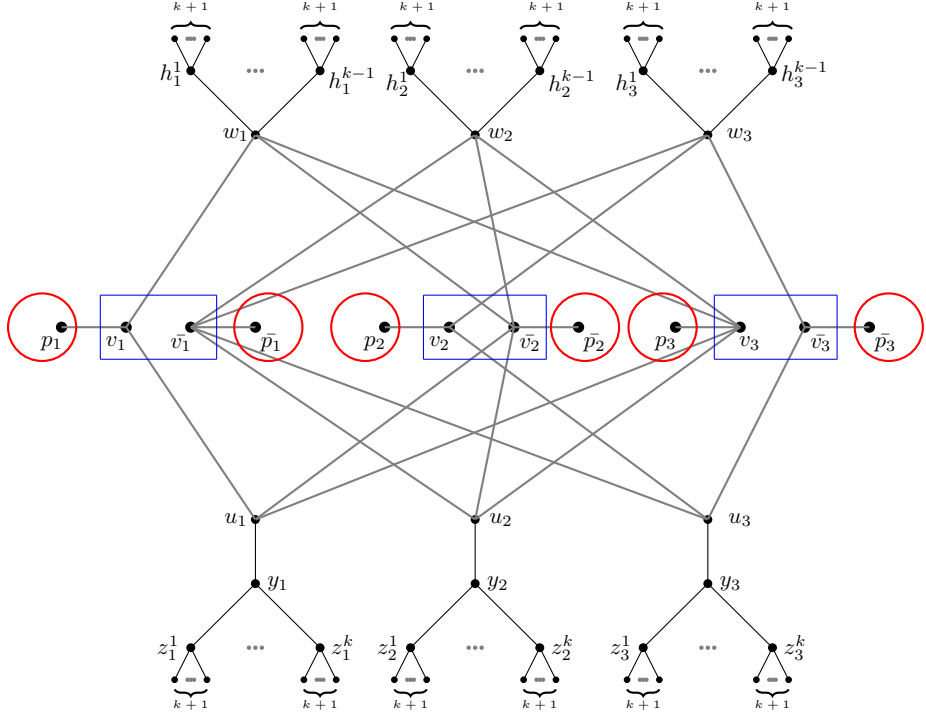

In this section, we study the problem complexity on bipartite graphs and obtain the following results. 
\begin{enumerate}
    \item Assuming ETH, there exists no algorithm that runs in time $\mathcal{O}^*(2^{o(n)})$ for bipartite graphs.
    \item The problem is NP-complete for $\Delta(G) \geq k+2$, when $k \geq 5$, even for bipartite graphs.
\end{enumerate} \vspace{3mm}
\textbf{Claim 1.} There always exists a \textit{minimum membership dominating set} for graphs with maximum degree, $\Delta(G) \leq k$. \vspace{2mm} \\
\textit{Proof.} We consider the following two cases based on the value of $\Delta(G)$. \vspace{3mm} \\ 
$\Delta(G) \leq$ $k-1$: The vertex set $V$ forms a \textit{minimum membership dominating set}. As each vertex $v \in V$ has at most $k$ vertices in its closed neighbourhood, the MMDS constraint is satisfied for each vertex $v \in V$. Therefore, it is always an YES-instance for this case. \vspace{2mm} \\
$\Delta(G) =$ $k$: We include all the vertices of degree at most $k-1$ to $S$. As all these vertices have at most $k$ vertices in their closed neighbourhood, they satisfy the MMDS constraint. We then find all the undominated vertices of $G$, let this set be denoted by $P$. We arbitrarily pick a vertex (say $u$) from $P$ and include it in $S$. We then update $P$ based on the recent addition of $u$ to $S$. We perform this until $P$ is empty. We add $u$ to $S$ only because it is undominated, so adding $u$ to $S$ will not violate MMDS constraint for any vertex of $G$. Hence, it is an YES-instance.\qed \vspace{2mm} \\
As the problem is linear-time solvable for maximum degree, $\Delta(G) \leq$ $k$, we study the problem complexity for higher values of $k$. \vspace{3mm} \\ 
From Theorem 2, we have that 3-CNF$^{\leq 3}$-XSAT problem can not be solved in $\mathcal{O}^*(2^{o(n)})$ time. We provide a reduction from 3-CNF$^{\leq 3}$-XSAT problem to show that \MMDS{} problem on bipartite graphs can not be solved in $\mathcal{O}^*(2^{o(n)})$ time. \vspace{2mm} \\
\textbf{Reduction.} \vspace{3mm} \\
Let $\phi$ be an instance of 3-CNF$^{\leq 3}$-XSAT problem with $n$ variables $x_1, x_2, $$..., x_n$ and $m$ clauses $C_1, C_2$,$..., C_m$ and each variable occurring in at most three clauses. We construct an instance of MMDS, $I = (G, k)$ as follows. 
\begin{itemize}
    \item For each clause $C_j$, we create three vertices $u_j$, $y_j$ and $w_j$. For each literal $x_i$, we create a vertex $v_i$ and for each literal $\bar{x_i}$, we create a vertex $\bar{v_i}$.
    \item If $x_i$ is part of a clause $C_j$, then $v_i$ is made adjacent to $u_j$ and $w_j$. Similarly, If $\bar{x_i}$ is part of a clause $C_j$, then $\bar{v_i}$ is made adjacent to $u_j$ and $w_j$.
    \item We make $u_j$ adjacent to $y_j$.
    \item For each vertex $w_j$, we create a vertex set $h_j^1, h_j^2, ..., h_j^{k-1}$ and we make $w_j$ adjacent to $h_j^1, h_j^2,...,h_j^{k-1}$.
    \item Each vertex of $h_j^1, h_j^2,...,h_j^{k-1}$ has $k+1$ pendant vertices adjacent to it.
    \item For each vertex $y_j$, we create a vertex set $z_j^1, z_j^2, ..., z_j^k$ and we make $y_j$ also adjacent to $z_j^1, z_j^2,...,z_j^k$.
    \item Each vertex among $z_j^1, z_j^2,...,z_j^k$ has $k+1$ pendant vertices adjacent to it.
    \item For each vertex $v_i$ (or $\bar{v_i}$), we create a vertex $p_i$ (or $\bar{p_i}$) and make both the vertices adjacent.
\end{itemize}
\textbf{Variable gadget.} For each vertex pair ($v_i$, $\bar{v_i}$), we create the vertices $b_i, c_i$ and $d_i$. We also create the sets $a_i^1, a_i^2,...,a_i^k$ and $f_i^1, f_i^2 ,..., f_i^{k-1}$. Vertices $v_i$ and $\bar{v_i}$ are made adjacent to $c_i$ and $d_i$. $d_i$ is adjacent to $f_i^1, f_i^2 ,..., f_i^{k-1}$. $c_i$ is adjacent to $b_i$. $b_i$ is also made adjacent to $a_i^1, a_i^2,...,a_i^k$. Each of $f_i^1, f_i^2 ,..., f_i^{k-1}$ and $a_i^1, a_i^2,...,a_i^k$ has $k+1$ pendant vertices connected to them. See \autoref{fig: Fig4} for an illustration. \vspace{3mm} \\
Note: Variable gadget is created only if both the vertices $v_i$ and $\bar{v_i}$ exists. If not, we simply avoid the creation of variable gadget for such vertex $v_i$ or $\bar{v_i}$. \vspace{3mm} \\
\textbf{Literal gadget.} For each literal $v_i$, we create the vertices $q_i$ and $r_i$. We also create a set $s_i^1, s_i^2,..., s_i^k$. $p_i$ is adjacent to vertex $q_i$ which is adjacent to vertex $r_i$. $r_i$ also has the vertices $s_i^1, s_i^2,..., s_i^k$ adjacent to it. Each of $s_i^1, s_i^2,...,s_i^k$ has $k+1$ pendant vertices connected to them. See \autoref{fig: Fig4} for an illustration. \vspace{2mm} \\ 
This concludes the construction of the reduced instance $I$. It is to be noted that, we obtain the reduced instance $G$, only after having the variable gadget and literal gadget at the respective places as highlighted in \autoref{fig: Fig7}. \vspace{2mm} \\
\setlength{\textfloatsep}{1\baselineskip plus 0\baselineskip minus 0\baselineskip}
%must be part of the solution%
\noindent \textbf{Lemma 1.} If $\phi$ has a satisfying assignment, then $I$ has a \textit{minimum membership dominating set} with membership value $k$.\vspace{2mm} \\
\textit{Proof.} Let $X = \{x_i | i \in [n]\} \rightarrow \{0, 1\}$ be a satisfying assignment for $\phi$. We obtain the corresponding \textit{minimum membership dominating set} as follows. 
\begin{enumerate}
    \item For each $i \in (1,2,...,n)$: \begin{itemize}
        \item Add $v_i$ to $S$ if $x_i$ = 1 in $X$; add $\bar{v_i}$ to $S$ if $x_i$ = 0 in $X$.
    % \item Add all the vertices $w_1, w_2, ..., w_{k-1}$; $z_1, z_2, ..., z_k$ and $x_1, x_2, ..., x_{k-1}$ from each gadget of $P$ to $S$.
        \item Add $s_i^1, s_i^2, ..., s_i^k$ and $p_i$ from literal gadget to $S$.
        \item Add $f_i^1, f_i^2, ..., f_i^{k-1}$ and $a_i^1,a_i^2,...,a_i^k$ from variable gadget to $S$.
        \end{itemize}
    \item Add the vertices $\bigcup_{j=1}^{j=m}h_j^1 \cup h_j^2 \cup ... \cup h_j^{k-1}$ to $S$.
    \item Add the vertices $\bigcup_{j=1}^{j=m}z_j^1 \cup z_j^2 \cup ... \cup z_j^k$ to $S$.
\end{enumerate} \vspace{2mm}
Consider the literal gadget. 
\begin{description}
    \item[$p \in p_i$ (or $\bar{p_i}$):] Vertex $p$ has one closed neighbour, the vertex itself in $S$. if $v_i$ is in $S$, then $p$ has two closed neighbours in $S$. Therefore, $p$ has at least one and at most two closed neighbours in $S$.
    \item[$p \in s_i^1, s_i^2, ..., s_i^k$:] Vertex $p$ has exactly one closed neighbour, the vertex itself in $S$ and no other neighbours are in $S$.
    \item[$p \in r_i$:] The $k$ neighbours of $p$: $s_i^1, s_i^2, ..., s_i^k$ are in $S$. The vertex $k$ has exactly $k$ closed neighbours in $S$.
    \item[$p \in q_i$:] Vertex $p$ has exactly one neighbour, $p_i$ in $S$. The vertex $p$ has exactly one closed neighbour in $S$.
\end{description} \vspace{2mm}
Consider the variable gadget. 
\begin{description}
    \item[$p \in \{f_i^1, f_i^2, ..., f_i^{k-1}\}:$] Vertex $p$ is in $S$ and no other neighbours of $p$ are in $S$. $p$ has exactly one closed neighbour in $S$.
    \item[$p \in d_i:$] Vertex $p$ has $k$ neighbours $f_i^1, f_i^2, ..., f_i^{k-1}$ in $S$ and one among $v_i$ and $\bar{v_i}$ in $S$. $p$ has exactly $k$ closed neighbours in $S$.
    \item[$p \in c_i:$] One neighbour of $p$, either $v_i$ or $\bar{v_i}$ in $S$. As no other neighbours of $p$ are in $S$, it has exactly one closed neighbour in $S$.
    \item[$p \in a_i^1, a_i^2, ..., a_i^k:$] Vertex $p$ is self dominated and no other neighbours of $p$ are in $S$. Therefore, $p$ has exactly one closed neighbour in $S$.
    \item[$p \in b_i:$] $p$ has exactly $k$ neighbours, $\{a_i^1, a_i^2, ..., a_i^k\}$ in $S$.
\end{description} \vspace{3mm} 
\begin{description}
    \item[$p \in z_j^1, z_j^2, ..., z_j^k:$] Vertex $p$ is self dominated and no other neighbour of $p$ is in $S$. Therefore, $p$ has exactly one closed neighbour in $S$.
    \item[$p \in y_j:$] Vertex $p$ has exactly $k$ closed neighbours, $\{z_j^1, z_j^2, ..., z_j^k\}$ in $S$.
    \item[$p \in u_j:$] $p$ has exactly one neighbour from its clause variables, the one which is set to true in $S$ and no other neighbours of $p$ are in $S$. Hence, $p$ has exactly one closed neighbour in $S$.
    \item[$p \in v_i$ (or) $\bar{v_i}$:] Vertex $p$ has its neighbour from the literal gadget, $p_i$ in $S$. If $p$ is a part of $S$, then it will have two closed neighbours in $S$. Otherwise, it will have exactly one closed neighbour in $S$.
    \item[$p \in w_j:$] Vertex $p$ has $k-1$ closed neighbours, $\{h_j^1, h_j^2, ..., h_j^{k-1}\}$ in $S$. It also has one neighbour from its clause variables in $S$. This leads to $p$ having exactly $k$ closed neighbours in $S$.
    \item[$p \in h_j^1, h_j^2, ..., h_j^{k-1}:$] Vertex $p$ is self dominated and no other neighbours of $p$ are in $S$. Therefore, $p$ has exactly one closed neighbour in $S$.
    \item [$p \in$] \textit{pendant vertex}: Vertex $p$ has exactly one vertex, its neighbour, in $S$. Hence, $p$ exactly one closed neighbour in $S$.
\end{description} \vspace{3mm}
As all the vertices of $S$ satisfy the MMDS constraint, we conclude that $I$ has a \textit{minimum membership dominating set} with membership value $k$. 
\qed \vspace{3mm} \\
\textbf{Lemma 2.} If $I$ has a \textit{minimum membership dominating set} with membership value $k$ then $\phi$ has a satisfying assignment.\vspace{2mm} \\
\textit{Proof.} 
\begin{itemize}
    \item The vertices with $k+1$ pendant vertices as neighbours must be a part of $S$. Therefore, the vertices $\{s_i^1, s_i^2, ..., s_i^k\}$; $\{f_i^1, f_i^2, ..., f_i^{k-1}\}$; $\{a_i^1, a_i^2, ..., a_i^k\}$; $\{z_j^1, z_j^2, ..., z_j^k\}$ and $\{h_j^1, h_j^2, ..., h_j^{k-1}\}$ are in $S$.
    \item The $k-1$ neighbours $\{f_i^1, f_i^2, ..., f_i^{k-1}\}$ of $d_i$ are in $S$. Therefore, at most one among $v_i$ and $\bar{v_i}$ can be $S$.
    \item The $k$ neighbours $\{a_i^1, a_i^2, ..., a_i^k\}$ of $b_i$ are in $S$, vertices $b_i$ and $c_i$ can not be a part of $S$. In order to dominate $c_i$, at least one among $v_i$ and $\bar{v_i}$ can be $S$. This leads to having exactly one vertex from $v_i$ and $\bar{v_i}$ in $S$.
    \item As the $k$ neighbours $\{s_i^1, s_i^2, ..., s_i^k\}$ of $r_i$ are in $S$, vertices $r_i$ and $q_i$ can not be a part of $S$. In order to dominate $q_i$; $p_i$ must be in $S$.
    \item With the vertices $\bigcup_{i=1}^{i=n} p_i$ (or $\bar{p_i}$) being a part of $S$, all the vertices $\bigcup_{i=1}^{i=n} (v_i \cup \bar{v_i})$ are dominated.
    \item As the $k-1$ neighbours $\{h_j^1, h_j^2, ..., h_j^{k-1}\}$ of $w_j$ are in $S$, at most one among the three vertices of clause $C_j$ are in $S$.
    \item Each vertex $y_j$ has $k$ neighbours, $\{z_j^1, z_j^2, ..., z_j^k\}$ in $S$. Hence, $y_j$ and $u_j$ cannot be a part of $S$.
    \item In order to dominate the vertex $u_j$, at least one among the three vertices of clause $C_j$ are in $S$. This leads to having exactly one vertex corresponding to the clause $C_j$ in $S$.
\end{itemize}
   The literals corresponding to these vertices forms a satisfying assignment. Hence, we conclude that $\phi$ has a satisfying assignment. \qed \vspace{3mm} \\
\noindent $G$ is a bipartite graph with bipartition, \\
\begin{center}
    $(\bigcup_{i=1}^n (p_i \cup \bar{p_i} \cup d_i \cup c_i \cup A_i \cup P_{F_i} \cup r_i \cup P_{S_i})  \cup P_H \cup W \cup U \cup Z)$
\end{center} and 
\begin{center}
    $(\bigcup_{i=1}^n (v_i \cup \bar{v_i} \cup q_i \cup P_{A_i} \cup F_i \cup b_i \cup S_i) \cup Y \cup H \cup P_Z),$ \vspace{3mm}
\end{center}  where $A_i$ = $\bigcup_{l=1}^{l=k} a_i^l$, $F_i$ = $\bigcup_{l=1}^{l=k-1} f_i^l$, $S_i$ = $\bigcup_{l=1}^{l=k} s_i^l$, $U$ = $\bigcup_{j=1}^{j=m} u_j$, $Y$ = $\bigcup_{j=1}^{j=m} y_j$, $H$ = $\bigcup_{j=1}^{j=m} \bigcup_{l=1}^{l=k-1} h_j^l$, $Z$ = $\bigcup_{j=1}^{j=m} \bigcup_{l=1}^{l=k} z_j^l$, $W$ = $\bigcup_{j=1}^{j=m} w_j$; $P_{A_i}, P_{F_i}, P_{S_i}, P_H$ and $P_Z$ represents the pendant vertices of the sets $A_i, F_i, S_i, H$ and $Z$ respectively. \vspace{2mm} \\
For each value of $i$, one vertex pair among $(p_i, v_i)$ and $(\bar{p_i}, \bar{v_i})$ is in $S$. Irrespective of the value of $k$, either $p_i$ or $\bar{p_i}$ has exactly two closed neighbours in $S$. Hence, $k \geq 2$. \vspace{3mm} \\
\textbf{Claim 2.} The size of the reduced instance of $I$ is linear in the size of the original instance $\phi$. \vspace{2mm} \\
\textit{Proof. } The size of the variable gadget shown in \autoref{fig: Fig4} is $\mathcal{O}(k^2)$. There will be at most $n$ of them. The size of the literal gadget shown in \autoref{fig: Fig4} is $\mathcal{O}(k^2)$. There will be at most $2n$ of them. The size of $I$ without both the gadgets, shown in \autoref{fig: Fig7} is $\mathcal{O}(m \cdot k^2)$. The size of the reduced instance $I$ is $|V| = \mathcal{O}(k^2(m+n))$. If $k=\mathcal{O}(1)$, $|V| = \mathcal{O}(m+n)$. \qed \vspace{3mm} \\
Hence, we have the following lower bound for the MMDS problem on bipartite graphs. \vspace{3mm} \\
\textbf{Theorem 5.}\textit{ There exists no algorithm that runs in time $\mathcal{O}^*(2^{o(n)})$ for the }\MMDS{} \textit{problem on bipartite graphs, for $k \geq 2$ (assuming that ETH holds).} \qed \vspace{3mm} \\
\textbf{Claim 3.} The lower bound on the maximum degree of the reduced instance is seven. \vspace{2mm} \\
\textit{Proof.} All the vertices of $G$, except $v_i$ and $\bar{v_i}$, have a maximum degree of $k+2$. \vspace{3mm} \\ A variable $x_i$ has the following two ways of appearing in $\phi$. \vspace{2mm} \\ \textit{Case 1:} \textit{Three times either in positive or negative form.} \vspace{2mm} \\ Let us assume that a variable $x_i$ occurrs three times in only positive form. This means that $\bar{x_i}$ does not exist in the formula and the variable gadget is not constructed. Vertex $v_i$ is adjacent to three neighbours in each of $\bigcup_{j=1}^m u_j$ and $\bigcup_{j=1}^m w_j$. It is also adjacent to $p_i$. So, the vertex $v_i$ has a degree of seven. \vspace{2mm} \\
\textit{Case 2:} \textit{One time in one form and two times in other form.} \vspace{2mm} \\
Let $x_i$ occurs two times and $\bar{x_i}$ occurs one time. Vertex $v_i$ is adjacent to two neighbours in each of $\bigcup_{j=1}^m u_j$ and $\bigcup_{j=1}^m w_j$. It is also adjacent to $p_i$. It also has two neighbours, $c_i$ and $d_i$ from the variable gadget. Vertex $\bar{v_i}$ is adjacent to one neighbour in each of $\bigcup_{j=1}^m u_j$ and $\bigcup_{j=1}^m w_j$. It is also adjacent to $p_i$. It also has two neighbours, $c_i$ and $d_i$ from the variable gadget. This leads to degree of seven and five for the vertices $v_i$ and $\bar{v_i}$, respectively. \vspace{2mm} \\
In both the cases, one among $v_i$ and $\bar{v_i}$ has a degree of seven, while the other vertex has a degree lesser than seven. Hence, we conclude that the lower bound on the maximum degree of $G$ is seven. \qed \vspace{3mm} \\ 
Irrespective of the value of $k$, one among the vertices $p_i$ and $\bar{p_i}$ has a degree of seven. Therefore, the maximum degree of the reduced instance $I$ is $k+2$ and it can not be less than seven. Hence, we arrive at the following theorem. \vspace{3mm} \\
\textbf{Theorem 6.} \textit{For $k\geq 5$, the }\MMDS{} \textit{problem with membership $k$ is NP-complete on bipartite graphs with maximum degree $k+2$}. \qed
\section{Parameterized complexity for structural parameters}
In this section, we study the parameterized complexity of the problem for 1) the twin cover and 2) the combined parameter distance to cluster, membership($k$). We prove that the problem is in FPT for both the cases.
\subsection{FPT parameterized by twin cover}
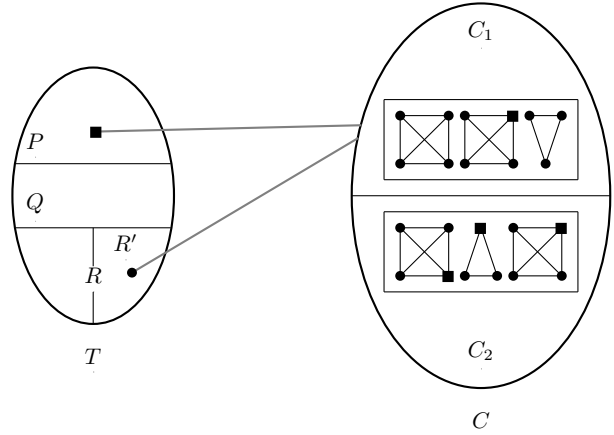
\begin{figure} [t]
\centering
    \begin{tikzpicture} [thick,scale=0.85, every node/.style={scale=0.85}]
        \draw (0, 0) ellipse (1.25 and 2);
        \filldraw (0, -2.75) circle (0cm) node[anchor=south]{$T$};
        \filldraw (-0.9, 0.6) circle (0cm) node[anchor=south]{$P$};
        \draw (6,0) ellipse (2 and 3);
        \draw[thin] (-1.2, 0.5) -- (1.2, 0.5);
        \filldraw (6, -3.75) circle (0cm) node[anchor=south]{$C$};
        \filldraw (-0.9, -0.4) circle (0cm) node[anchor=south]{$Q$};
        \draw[thin] (-1.2, -0.5) -- (1.2, -0.5);
        \filldraw (0, -1.5) circle (0cm) node[anchor=south]{$R$};
        \draw[thin] (0, -0.5) -- (0, -1.1);
        \draw[thin] (0, -2) -- (0, -1.5);
        \filldraw (0.5, -1) circle (0cm) node[anchor=south]{$R'$};
        \draw[thin] (4, 0) -- (8, 0);
        \filldraw (6, 2.3) circle (0cm) node[anchor=south]{$C_1$};
        \filldraw (6, -2.7) circle (0cm) node[anchor=south]{$C_2$};

        \draw[gray] (0, 1) -- (4.15, 1.1);
        \draw[gray] (0.6, -1.2) -- (4.1, 0.9);
        
        \filldraw (-0.03, 0.92) rectangle ++(4pt, 4pt);
        \filldraw (0.6, -1.2) circle (0.06cm);

        \draw[thin] (4.5, -0.25) -- (7.5, -0.25);
        \draw[thin] (4.5, -1.5) -- (7.5, -1.5);
        \draw[thin] (4.5, -1.5) -- (4.5, -0.25);
        \draw[thin] (7.5, -0.25) -- (7.5, -1.5);

        \draw[thin] (4.75, -0.5) -- (5.5, -0.5);
        \draw[thin] (4.75, -0.5) -- (5.5, -1.25);
        \draw[thin] (4.75, -1.25) -- (5.5, -1.25);
        \draw[thin] (4.75, -1.25) -- (4.75, -0.5);
        \draw[thin] (4.75, -1.25) -- (5.5, -0.5);
        \draw[thin] (5.5, -0.5) -- (5.5, -1.25);
        \filldraw (4.75, -0.5) circle (0.06cm);
        \filldraw (5.5, -0.5) circle (0.06cm);
        \filldraw (4.75, -1.25) circle (0.06cm);
        \filldraw (5.42, -1.33) rectangle ++(4pt, 4pt);

        \draw[thin] (5.75, -1.25) -- (6.25, -1.25);
        \draw[thin] (5.75, -1.25) -- (6, -0.5);
        \draw[thin] (6.25, -1.25) -- (6, -0.5);
        \filldraw (5.75, -1.25) circle (0.06cm);
        \filldraw (6.25, -1.25) circle (0.06cm);
        \filldraw (5.92, -0.58) rectangle ++(4pt, 4pt);

        \draw[thin] (6.5, -0.5) -- (7.25, -0.5);
        \draw[thin] (6.5, -1.25) -- (7.25, -1.25);
        \draw[thin] (6.5, -1.25) -- (7.25, -0.5);
        \draw[thin] (6.5, -0.5) -- (7.25, -1.25);
        \draw[thin] (6.5, -1.25) -- (6.5, -0.5);
        \draw[thin] (7.25, -0.5) -- (7.25, -1.25);
        \filldraw (6.5, -0.5) circle (0.06cm);
        \filldraw (6.5, -1.25) circle (0.06cm);
        \filldraw (7.25, -1.25) circle (0.06cm);
        \filldraw (7.17, -0.58) rectangle ++(4pt, 4pt);

        \draw[thin] (4.5, 0.25) -- (7.5, 0.25);
        \draw[thin] (4.5, 1.5) -- (7.5, 1.5);
        \draw[thin] (4.5, 1.5) -- (4.5, 0.25);
        \draw[thin] (7.5, 0.25) -- (7.5, 1.5);

        \draw[thin] (4.75, 0.5) -- (5.5, 0.5);
        \draw[thin] (4.75, 0.5) -- (5.5, 1.25);
        \draw[thin] (4.75, 1.25) -- (5.5, 0.5);
        \draw[thin] (4.75, 1.25) -- (5.5, 1.25);
        \draw[thin] (4.75, 1.25) -- (4.75, 0.5);
        \draw[thin] (5.5, 0.5) -- (5.5, 1.25);
        \filldraw (4.75, 0.5) circle (0.06cm);
        \filldraw (5.5, 0.5) circle (0.06cm);
        \filldraw (4.75, 1.25) circle (0.06cm);
        \filldraw (5.5, 1.25) circle (0.06cm);

        \draw[thin] (6.75, 1.25) -- (7.25, 1.25);
        \draw[thin] (6.75, 1.25) -- (7, 0.5);
        \draw[thin] (7.25, 1.25) -- (7, 0.5);
        \filldraw (6.75, 1.25) circle (0.06cm);
        \filldraw (7.25, 1.25) circle (0.06cm);
        \filldraw (7, 0.5) circle (0.06cm);

        \draw[thin] (5.75, 0.5) -- (6.5, 0.5);
        \draw[thin] (5.75, 0.5) -- (6.5, 1.25);
        \draw[thin] (5.75, 1.25) -- (6.5, 0.5);
        \draw[thin] (5.75, 1.25) -- (6.5, 1.25);
        \draw[thin] (5.75, 1.25) -- (5.75, 0.5);
        \draw[thin] (6.5, 0.5) -- (6.5, 1.25);
        \filldraw (5.75, 0.5) circle (0.06cm);
        \filldraw (5.75, 1.25) circle (0.06cm);
        \filldraw (6.5, 0.5) circle (0.06cm);
        \filldraw (6.42, 1.17) rectangle ++(4pt, 4pt);
    \end{tikzpicture}
    \caption{Partitioning of the vertex set $V$ into sets $T$ and $C$, where $T$ is the twin cover and $C$ is the union of clique sets outside $T$. Square shaped vertices are a part of $S$.}
    \label{fig: Fig8}
\end{figure}

\noindent We consider the structural parameter twin cover. With the help of some crucial observations, we show that the problem is in FPT. The parameter twin cover is defined as follows. \vspace{3mm} \\
\textbf{Definition 1.} For a graph $G = (V, E)$, the parameter \textit{twin cover} is the cardinality of the smallest set $T \subseteq V$ such that every component in $V\setminus T$ is a set of true twins in $G$.\vspace{2mm} \\ 
The parameter twin cover of size at most $k$ (if exists), can be obtained in FPT time, in the following way. \vspace{3mm} \\
\textbf{Theorem 7.}~\cite[Theorem 4.]{GNN} \textit{If a minimum twin cover in $G$ has size at most $k$, then it is possible to compute a twin cover of size at most $k$ in time $\mathcal{O}(|E|+k|V|+1.2738^k)$.} \qed \vspace{2mm} \\
Given a graph $G$, we partition the vertex set $V$ into sets $T$ and $C$, where $T$ represents the twin cover and $C$ is the union of clique sets outside $T$. A clique set is a union of cliques with the same adjacency in $T$. We guess the set $P = T \cap S$, obtain $Q = N(P) \cap T$ and $R$ = $T \setminus \{P \cup Q\}$. The guess of $P$ can be made in $2^{|T|}$ ways. We check whether any vertex has more than $k$ closed neighbours in $P$; if yes, we discard the guess. Let $C_1$ denote the union of clique sets with at least one neighbour in $P$ and $C_2$ denote the union of clique sets with no neighbor in $P$. See \autoref{fig: Fig8} for an illustration. \vspace{3mm} \\ \indent The vertices in $C_2$ must dominate themselves, so one vertex from each clique for all such clique sets must belong to $S$. We select one vertex arbitrarily to be a part of $S$ from each clique for all these clique sets. As all the vertices of the clique have the same neighbourhood, picking any one vertex would work. Let these vertices from $C_2$ be represented by $X$.  At this point, we again verify whether any vertex has more than $k$ closed neighbors in $S$. \vspace{3mm} \\
\setlength{\textfloatsep}{1\baselineskip plus 0\baselineskip minus 0\baselineskip}
\indent Currently, all the vertices of $P, Q$, and $C$ are dominated and a subset of vertices of $R$ are dominated. Obtain the vertices from $R$ that are not yet dominated. Let these set of vertices be denoted by $R'$. Note that the vertices of $R'$ do not have any neighbor in $S$. The only way to dominate these vertices is by having their neighbours from $C_1$ in $S$. If we choose to add vertices from a clique set $\mathbb{C} \in C_1$ to $S$, then adding exactly one vertex from $\mathbb{C}$ would be enough to meet our needs. We guess the clique sets from $C_1$ to pick a vertex from, to make it a part of $S$. This guess can be made in at most $2^{2^{|T|}}$ ways. Let $Y$ denote the set of vertices of this guess from $C_1$. We verify whether $P \cup X \cup Y$ forms a \textit{minimum membership dominating set}, in polynomial time. The problem can be solved in $O^*(2^{|T|} \cdot 2^{2^{|T|}})$ time, where $T$ is the twin cover size. This sums up the proof of the following theorem. \vspace{3mm} \\
\textbf{Theorem 8.} \textit{Given a graph $G = (V, E)$, $T \subseteq V$ is a twin cover of $G$, the }\MMDS{}\textit{ problem can be solved in $\mathcal{O}^*(2^{|T|} \cdot 2^{2^{|T|}})$ time.} \qed
\subsection{FPT parameterized by the combined parameter distance to cluster, membership($k$) }
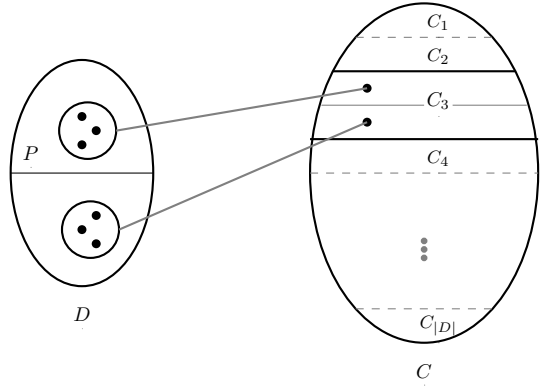
\begin{figure} [t]
\centering
    \begin{tikzpicture} [thick,scale=0.75, every node/.style={scale=0.75}]
        \draw (12, 0) ellipse (1.25 and 2);
        \filldraw (12, -2.75) circle (0cm) node[anchor=south]{$D$};
        \filldraw (11.1, 0.1) circle (0cm) node[anchor=south]{$P$};
        \draw (18,0) ellipse (2 and 3);
        \draw[thin] (10.75, 0) -- (13.25, 0);
        \filldraw (18, -3.75) circle (0cm) node[anchor=south]{$C$};
        \draw[thin, gray] (16.2, 1.2) -- (18, 1.2);
        \draw[thin, gray] (18.5, 1.2) -- (19.8, 1.2);
        \draw[thick] (16, 0.6) -- (20, 0.6);
        \draw[thin, gray, dashed] (16.8, 2.4) -- (19.2, 2.4);
        \draw[thick] (16.4, 1.8) -- (19.6, 1.8);
        \draw[thin, gray, dashed] (16, 0) -- (20, 0);
        \draw[thin, gray, dashed] (16.8, -2.4) -- (19.2, -2.4);
        \filldraw (18.25, 2.4) circle (0cm) node[anchor=south]{\scriptsize{$C_1$}};
        \filldraw (18.25, 1.8) circle (0cm) node[anchor=south]{\scriptsize{$C_2$}};
        \filldraw (18.25, 1) circle (0cm) node[anchor=south]{$C_3$};
        \filldraw (18.25, 0) circle (0cm) node[anchor=south]{\scriptsize{$C_4$}};
        \filldraw (18.25, -3) circle (0cm) node[anchor=south]{\scriptsize{$C_{|D|}$}};

        \filldraw (12, 1) circle (0.06cm);
        \filldraw (12.25, 0.75) circle (0.06cm);
        \filldraw (12, 0.5) circle (0.06cm);

        \filldraw (12.25, -0.75) circle (0.06cm);
        \filldraw (12, -1) circle (0.06cm);
        \filldraw (12.25, -1.25) circle (0.06cm);

        \draw (12.1, 0.75) ellipse (0.5 and 0.5);
        \draw (12.15, -1) ellipse (0.5 and 0.5);

        \filldraw (17, 1.5) circle (0.06cm);
        \filldraw (17, 0.9) circle (0.06cm);
        
        \draw[gray] (17, 1.5) -- (12.6, 0.75);
        \draw[gray] (17, 0.9) -- (12.65, -1);

        \filldraw[gray] (18, -1.35) circle (0.04cm);
        \filldraw[gray] (18, -1.5) circle (0.04cm);
        \filldraw[gray] (18, -1.2) circle (0.04cm);
        
    \end{tikzpicture}
    \caption{Partitioning of the vertex set $V$ into sets $D$ and $C$, where $|D|$ is the distance to cluster and $C$ is the union of clique sets outside $D$. The collective neighbourhood of $C_3$ in $D$ is shown.}
    \label{fig: Fig9}
\end{figure}

We consider the combined parameter distance to cluster, membership($k$). We propose an exponential kernel, which also infers that the problem is in FPT. The parameter distance to cluster is defined as follows.\vspace{2mm} \\
\textbf{Definition 2.} For a graph $G = (V, E)$, the parameter \textit{distance to cluster} is the cardinality of the smallest set $D \subseteq V$ such that every component in $V\setminus D$ is a clique.\vspace{3mm} \\
\noindent Consider a graph $G$, partitioned into $D$ and $C$ with $|D|$ as the distance to cluster and $C$ as the union of clique sets outside $D$. The collective neighbourhood of a set $X$ in another set $Y$ is the set of vertices in $N[X] \cap Y$. In this case, clique set represents a union of cliques such that each clique in a clique set has the same collective neighbourhood in $D$. We guess the subset of vertices $P = S \cap D$ and compute $Q = S \cap C$. We have $2^{|D|}$ unique subset of vertices from $D$. We use $C_i$ to denote the union of clique sets with $i$ vertices in the collective neighbourhood in $D$. See \autoref{fig: Fig9} for an illustration. \vspace{3mm} \\
For the rest of the section, we use $\mathbb{C}$ to denote an arbitrary clique set from $C_i$. \vspace{3mm} \\
Consider a clique $C'$ from $\mathbb{C}$ in which vertices $u$ and $v$ have the same adjacency in $D$. \vspace{3mm} \\
\textbf{Reduction Rule 1.} \textit{If $u$ and $v$ are true twins, then delete either $u$ or $v$.} \vspace{3mm} \\ 
\textbf{Lemma 3.} \textit{Reduction Rule 1 is safe}. \vspace{2mm} \\
\textit{Proof.} As $u$ and $v$ are true twins, having one of them in the solution instead of both, does not alter the outcome of the MMDS constraint for any vertex in the graph. So, we delete one of the two vertices from the clique arbitrarily. \qed \vspace{3mm} \\
\noindent \textbf{Lemma 4.} \textit{The size of the largest clique in $\mathbb{C}$ is at most $2^i$.} \vspace{2mm} \\
\textit{Proof.} We repeatedly apply the reduction rule 1 on each pair of true twins of $C'$, and we finally end up with at most $2^i$ vertices that have a unique neighbourhood in $D$. \qed \vspace{3mm} \\
The graph obtained from $G$, after repeatedly applying reduction rule 1, is denoted by $G'$. \vspace{3mm} \\ 
\noindent We classify the cliques of $C$ into two types. Type $A$: cliques that have at least one vertex not adjacent to any vertex in $P$. Type $B$: cliques that have all the vertices adjacent to a vertex in $P$. \vspace{3mm} \\
\noindent First, we consider the cliques from $\mathbb{C}$ of Type $A$ and bound them. \vspace{3mm} \\
\noindent \textbf{Lemma 5.} \textit{There are at most $i \cdot k$ cliques in $\mathbb{C}$ of Type A}. \vspace{2mm} \\
\textit{Proof.} Consider a clique $C'$ from $\mathbb{C}$ that has at least one vertex (say $v$) not adjacent to any vertex in $P$. $v$ has to be dominated by some vertex of $C'$. So, $S$ consists of at least one vertex from each of these cliques. There are $i$ vertices in $D$ that correspond to the clique set. For all these $i$ vertices to satisfy the MMDS constraint, each of these vertices can only be adjacent to at most $k$ such cliques. We can have a total of at most $i \cdot k$ cliques of such sort. If there are more than $i \cdot k$ cliques, we simply reject the guess. \qed \vspace{2mm} \\
\noindent Now, we consider the cliques from $\mathbb{C}$ of Type $B$ and bound them. \vspace{3mm} \\
\textbf{Reduction Rule 2.} If there are more than $2^i$ cliques in $\mathbb{C}$ of Type $B$ with the same adjacency in $D$ then delete the cliques until only $2^i$ cliques in $\mathbb{C}$ of Type $B$ with the same adjacency in $D$ remains. \vspace{3mm} \\
\noindent \textbf{Lemma 6.} \textit{Reduction Rule 2 is safe.} \vspace{2mm} \\
\textit{Proof.} Consider a set of cliques from $\mathbb{C}$ of Type $B$ with the same adjacency in $D$. In the worst-case scenario, we might need all the vertices of a clique to dominate the vertices in $D$. We might end up picking $2^i$ vertices from $2^i$ unique cliques (one unique corresponding vertex from each clique). Having any more than $2^i$ cliques with the same adjacency in $D$ does not alter the solution. Hence, they can be simply discarded and we end up having at most $2^i$ cliques of Type $B$ with the same adjacency in $D$. \qed \vspace{3mm} \\
After applying the Reduction Rule 2 on $G'$, we obtain the following lemma. \vspace{3mm} \\ 
\noindent \textbf{Lemma 7.} \textit{There are at most $2^i$ cliques in $\mathbb{C}$ of Type $B$ with the same adjacency in $D$.} \qed \vspace{3mm} \\
The graph obtained from $G'$, after applying reduction rule 2, is denoted by $G''$. \vspace{3mm} \\ 
\textbf{Reduction Rule 3.} If there are more than $2^{2^{i^{i}}}$ cliques in $\mathbb{C}$ of Type $B$ with the unique neighbourhood in $D$ then delete the cliques until only $2^{2^{i^{i}}}$ cliques in $\mathbb{C}$ of Type $B$ with the unique neighbourhood in $D$ remains. \vspace{3mm} \\
\noindent \textbf{Lemma 8.} \textit{Reduction Rule 3 is safe.} \vspace{3mm} \\
\textit{Proof.} Consider a set $D'$ from $D$ that corresponds to $\mathbb{C}$. The size of a clique in $\mathbb{C}$ is bounded by $2^i$. We consider the presence of all possible cliques in $\mathbb{C}$ with the unique neighbourhood in $D'$. So, for each vertex in $D'$, we have at most $2^{2^{i}}$ ways of forming the adjacency with the clique. This is possible for all the $i$ vertices. This leads to a maximum of $2^{2^{i^{i}}}$ ways of forming the adjacency between the vertices of $D'$ and cliques in $\mathbb{C}$. This gives us $2^{2^{i^{i}}}$ unique cliques. The total number of cliques from $\mathbb{C}$ of Type $B$ with the unique neighbourhood in $D$ is $2^{2^{i^{i}}}$. Having any more than $2^{2^{i^{i}}}$ cliques with the unique neighbourhood in $D$ does not alter the solution. Hence, they can be simply discarded and we end up having at most $2^{2^{i^{i}}}$ cliques of Type $B$ with the unique neighbourhood in $D$. \qed \vspace{3mm} \\
After applying the Reduction Rule 3 on $G''$, we obtain the following lemma. \vspace{3mm} \\ 
\noindent \textbf{Lemma 9.} \textit{There are at most $2^{2^{i^{i}}}$ cliques in $\mathbb{C}$ of Type $B$ with the unique neighbourhood in $D$}. \qed \vspace{3mm} \\
\noindent \textbf{Lemma 10.} \textit{Given a graph $G = (V, E)$, $D \subseteq V$ is a distance to cluster of $G$, there exists an exponential kernel of size at most $ |D| + |D| \cdot$ $4^{|D|} \cdot (|D| \cdot k$  $+ 2^{2^{|D|^{|D|}}} \cdot 2^{|D|})$.} \vspace{2mm} \\
\textit{Proof.} 
\begin{itemize}
    \item From Lemma 4, we have a bound of $2^i$ on the size of the clique outside the distance to cluster. 
    \item From Lemma 5, we have a bound of $i \cdot k$ on the number of cliques of Type $A$. 
    \item From Lemma 7, we have a bound of $2^i$ on the number of cliques of Type $B$, with the same adjacency in $D$.
    \item From Lemma 9, we have a bound of $2^{2^{i^i}}$ on the number of cliques of Type $B$, with the unique neighbourhood in $D$.
\end{itemize} Combining Lemma 5, Lemma 7 and Lemma 9, we have a bound of $(i \cdot k$ + $ 2^{2^{i^{i}}} \cdot 2^i)$ on the number of cliques in $\mathbb{C}$. The size of $\mathbb{C}$ will be at most $2^i \cdot (i \cdot k$ + $ 2^{2^{i^{i}}} \cdot 2^i)$. There will be at most $\binom{|D|} {i}$ clique sets in $C_i$. The size of $C_i$ is upper bounded by $\binom{|D|} {i} \cdot 2^i \cdot (i \cdot k$ + $ 2^{2^{i^{i}}} \cdot 2^i)$. The value of $i$ varies from 1 to $|D|$. Hence, we obtain the following upper bound on the kernel size. \vspace{3mm} \\
    $=|D| + \sum_{i=1}^{|D|}$ $ \binom{|D|} {i} \cdot 2^i \cdot (i \cdot k$ + $ 2^{2^{i^{i}}} \cdot 2^i)$ \vspace{1mm} \\
    $=\mathcal{O} (|D| + |D| \cdot$ $2^{|D|} \cdot  2^{|D|} \cdot (|D| \cdot k$ + $ 2^{2^{|D|^{|D|}}} \cdot 2^{|D|})) $. \vspace{1mm} \\
    $=\mathcal{O} ( |D| + |D| \cdot$ $4^{|D|} \cdot (|D| \cdot k$ +  $ 2^{2^{|D|^{|D|}}} \cdot 2^{|D|}))$.\qed \vspace{2mm} \\
\noindent As there exists an exponential kernel for the problem in terms of the parameters distance to cluster, membership($k$); we obtain the following result.\vspace{2mm} \\ 
\textbf{Theorem 9.} \textit{Given a graph $G = (V, E)$, $D \subseteq V$ such that $|D|$ is distance to cluster of $G$, the }\MMDS{}\textit{ problem is in FPT parameterized by distance to cluster, membership($k$).} \qed
\section{Linear time algorithm for trees}
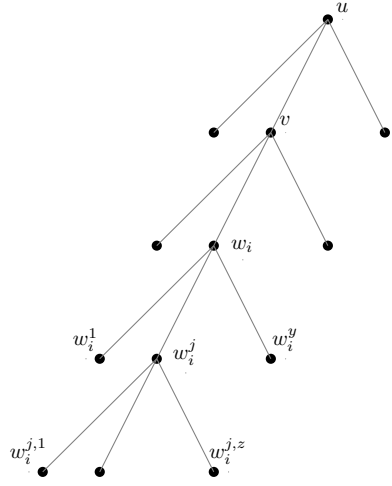
\begin{figure}
    \centering
    \begin{tikzpicture} [thick,scale=0.75, every node/.style={scale=0.8}]
        \filldraw (20,20) circle (2pt);
        \filldraw (18,18) circle (2pt);
        \filldraw (19,18) circle (2pt);
        \filldraw (21,18) circle (2pt);
        \filldraw (17,16) circle (2pt);
        \filldraw (18,16) circle (2pt);
        \filldraw (20,16) circle (2pt);
        \filldraw (16,14) circle (2pt);
        \filldraw (17,14) circle (2pt);
        \filldraw (19,14) circle (2pt);
        \filldraw (15,12) circle (2pt);
        \filldraw (16,12) circle (2pt);
        \filldraw (18,12) circle (2pt);
        \filldraw (15,12) circle (0pt);
        \filldraw (15,11) circle (0pt);

        \draw[thin, gray] (20, 20) -- (18, 18);
        \draw[thin, gray] (20, 20) -- (19, 18);
        \draw[thin, gray] (20, 20) -- (21, 18); 

        \draw[thin, gray] (19, 18) -- (17, 16);
        \draw[thin, gray] (19, 18) -- (18, 16);
        \draw[thin, gray] (19, 18) -- (20, 16); 

        \draw[thin, gray] (18, 16) -- (16, 14);
        \draw[thin, gray] (18, 16) -- (17, 14);
        \draw[thin, gray] (18, 16) -- (19, 14); 

        \draw[thin, gray] (17, 14) -- (15, 12);
        \draw[thin, gray] (17, 14) -- (16, 12);
        \draw[thin, gray] (17, 14) -- (18, 12); 
        
        \filldraw (20.25, 20) circle (0cm) node[anchor=south]{$u$};
        \filldraw (19.25, 18) circle (0cm) node[anchor=south]{$v$};
        \filldraw (18.5, 15.75) circle (0cm) node[anchor=south]{$w_i$};
        \filldraw (15.75, 14) circle (0cm) node[anchor=south]{$w_i^1$};
        \filldraw (17.5, 13.75) circle (0cm) node[anchor=south]{$w_i^j$};
        \filldraw (19.25, 14) circle (0cm) node[anchor=south]{$w_i^y$};
        \filldraw (14.75, 12) circle (0cm) node[anchor=south]{$w_i^{j, 1}$};
        \filldraw (18.25, 12) circle (0cm) node[anchor=south]{$w_i^{j, z}$};
        
    \end{tikzpicture}
    \caption{Handling the scenario of $M(u)$ = 1 when $M(v)$ = 0.}
    \label{fig: Fig10}
\end{figure}

We study the complexity of the MMDS problem for trees. We use the dynamic programming approach to compute the partial solutions at each node, starting from leaves and reaching all the way up to the root node.   This gives us a linear-time algorithm for the problem. \vspace{3mm} \\
Consider a tree, $G = (V, E)$ rooted at an arbitrary node $r$. $G_u$ indicates the induced subgraph containing $u$ and all its descendants. We use $S_u$ to denote a \textit{minimum membership dominating set} of $G_u$. At each node, we store four boolean variables, that are $M^+(u), M^-(u), M(u)$ and $M^{'}(u)$. For any given node $u$, we define the variables $M^+(u), M^-(u), M(u)$ and $M^{'}(u)$ in the following way. 
\begin{itemize}
    \item $M^+(u)$ represents whether there exists a \textit{minimum membership dominating set} for $G_u$ with $u$ in $S_u$.
    \item $M^-(u)$ represents whether there exists a \textit{minimum membership dominating set} for $G_u$ with $u$ not in $S_u$.
    \item $M(u)$ indicates whether there exists a \textit{minimum membership dominating set} for $G_u$.
    \item $M^{'}(u)$ indicates whether $M^-(u) = 0$ only because $u$ is undominated.
\end{itemize}
We obtain the values of $M^+(u)$, $M^-(u), M(u)$ and $M^{'}(u)$ as follows:
\begin{itemize}
    \item $M^+(u) =
    \begin{cases}
        1, &\text{if there exists an $S_u$ with $u$ a part of $S$.} \\
        0, & \text{otherwise}
    \end{cases} 
    $ \vspace{2mm}
    \item $M^-(u) =
    \begin{cases}
        1, &\text{if there exists an $S_u$ with $u$ not a part of $S$.} \\
        0, & \text{otherwise}
    \end{cases}
    $ \vspace{2mm}
    \item $M(u)$ = max($M^+(u), M^-(u)$). \vspace{2mm}
    \item $M^{'}(u) =
    \begin{cases}
        1, &\text{if (1) For the subgraph $G_u$, $M^-(u) = 0$ and} \\ &\text{(2) There exist an $S_u$ such that every node in $G_u$ except $u$} \\ &\text{satisfies the MMDS constraint and $u$ is undominated.} \\
        0, & \text{otherwise}
    \end{cases}
    $ \vspace{2mm}
\end{itemize}
The existence of a \textit{minimum membership dominating set} for $G$ can be obtained based on the value of $M(r)$. \vspace{3mm} \\
\textbf{Leaf node:} \vspace{2mm} \\
For a leaf node $u$, if $u$ is part of $S$ then $M^+(u)$ is set to 1. If $u$ is not part of $S$ then $M^-(u)$ is set to 0. As $u$ do not have a child node, $M^{'}(u)$ is set to 1. \vspace{2mm} \\
$M^+(u) = 1, M^-(u) = 0, M(u) = 1$ and $M^{'}(u) = 1$. \vspace{3mm} \\ 
\textbf{Non-leaf node: }  \vspace{2mm} \\
For a non-leaf node $u$, we use set $C(u)$ at each node, to store the child nodes of $u$. We define three boolean variables $P(u), Q(u)$ and $R(u)$. $P(u)$ and $Q(u)$ indicates whether there exists a \textit{minimum membership dominating set}, with at most $k$ of its child nodes in $S$ and with at most $k-1$ of its child nodes in $S$, respectively. $R(u)$ indicates whether there exists a \textit{minimum membership dominating set} with no child of $u$ (say $v$) having $k$ closed neighbours in $S_v$. The values of $P(u), Q(u)$ and $R(u)$ can be computed as follows. \vspace{3mm} \\
\begin{itemize}
    \item 
        $P(u)= 
    \begin{cases}
        0,              & \text{if } \sum_{\substack{v \in C(u) \\ M^-(v) = 0}} (1-M^{'}(v)) > k\\
        1,              & \text{otherwise}
    \end{cases}
    $ \vspace{2mm}
    \item 
        $Q(u)= 
    \begin{cases}
        0,              & \text{if } \sum_{\substack{v \in C(u) \\ M^-(v) = 0}} (1-M^{'}(v)) > k-1\\
        1,              & \text{otherwise}
    \end{cases}
    $ \vspace{2mm}
    \item 
        $R(u)= 
    \begin{cases}
        0,              & \text{if } ((1-M^-(u))(1-M^{'}(u)) + \sum_{\substack{v \in C(u) \\ M^-(v) = 0}} (1-M^{'}(v))) > k-1\\
        1,              & \text{otherwise}
    \end{cases}
    $ \vspace{2mm}
\end{itemize}
We compute the values of $M^-(u), M^{'}(u), M^+(u)$ and $M(u)$ as follows. \vspace{3mm} \\
\textbf{Computing $M^-(u)$:} \vspace{3mm} \\
    % We check whether there exists \textit{an $S_u$ such that $u$ has at most $k$ neighbours in $S_u$}. For each child node (say $v$) of $u$, we verify whether there exists an $S_v$. We check whether for at least one child node (say $v$) of $u$, $M^{'}(v) = 1$\vspace{3mm} \\
We compute the value of $M^-(u)$ as follows. \vspace{2mm} \\
$M^-(u) = (\prod_{v \in C(u)} M(v)) \cdot (\vee_{\substack{v \in C(u) \\ M^-(v) = 0}} (1-M^{'}(v))) \cdot P(u)$ \vspace{3mm} \\ 
Here, $M^-(u)$ is 1, if and only if, 
\begin{itemize}
    \item For each child node (say $v$) of $u$, $M(v) = 1$.
    \item Among all the child nodes (say $v$) of $u$, $M^{'}(v)$ is 0 for at least one of them. This will make sure that $u$ is dominated.
    \item Among all the child nodes (say $v$) of $u$, $M^{'}(v)$ is 0 for at most $k$ of them.
\end{itemize}\vspace{5mm}
\textbf{Computing $M^{'}(u)$:} \vspace{3mm} \\
% We check whether there exists \textit{an $S_u$ with $u$ not a part of $S_u$}. We obtain whether $u$ is undominated and for each child node (say $v$) of $u$, there is an $S_v$ with $v$ not a part of $S_v$. \vspace{3mm} \\
We compute the value of $M^{'}(u)$ as follows. \vspace{2mm} \\
$M^{'}(u) = (1-M^-(u)) \cdot (\prod_{v \in C(u)} M^-(v))$. \vspace{3mm} \\
Here, $M^{'}(u)$ is 1, if and only if, 
\begin{itemize}
    \item $M^-(u) = 0$.
    \item For each child node (say $v$) of $u$, $M^-(v)$ is 1.
\end{itemize} \vspace{5mm}
\textbf{Computing $M^+(u)$:} \vspace{3mm} \\
    We check whether there exists \textit{an $S_u$ such that $u$ has at most $k-1$ neighbours in $S_u$}. For each child node (say $v$) of $u$, we verify whether there exists an $S_v$ such that $v$ has at most $k-1$ closed neighbours in $S_v$. \vspace{2mm} \\
Consider a node $u$, with a child node $v$ such that $M(v) = 0$. This clearly indicates that no \textit{minimum membership dominating set} exists for $G_v$. But, this can still yield $M(u) = 1$, and we will also handle this scenario. \vspace{3mm} \\
The only reason for which $M(u) = 1$ even after $M(v) = 0$ for the child node $v$ is if $M(v) = 0$ only because $M^{'}(v) = 1$. $u$ being a part of $S_u$, can dominate $v$. As $u$ can only dominate $v$, it is to be noted that all the other MMDS constraints for all the nodes of $G_v$ must be satisfied. To support this, we have the following conditions.
\begin{enumerate}
    \item As $v$ is not a part of $S_v$; $S_{w_i}$ must exist for each child node $w_i$ of $v$. Hence, for each child node $w_i$ of $v$, $M(w_i) = 1$.
    \item In order for $v$ to be undominated, $v$ must not be a part of $S_v$ and all the child nodes of $v$ must not be a part of $S_v$.
    \begin{itemize}
        \item \textit{$v$ to be not a part of $S_v$:} for at least one child node of $v$ (say $w_i$), exactly $k$ child nodes of $w_i$ must be in $S_v$.
        \item \textit{No child node of $v$ to be a part of $S_v$:} for each child node of $v$ (say $w_i$), either $k$ child nodes of $w_i$ must be in $S_v$ or there must exist some child node of $w_i$ (say $w_i^j$) such that $k$ nodes from $\{w_i^j$ and child nodes of $w_i^j\}$, must be in $S_v$. See \autoref{fig: Fig10} for an illustration.
    \end{itemize}
    For each node $w_i \in C(v)$, one of the following two must hold: \begin{itemize}
        \item $M^{'}(x) = 0$ for $k$ nodes from $x \in \{w_i^1, w_i^2,...,w_i^y\}$, where $w_i^j$ represents a child node of $w_i$. \textbf{Note:} This condition must hold for at least one node of $w_i \in C(v)$.
        \item Let $w_i^{j, l}$ represents a child node of $w_i^j$. $M^{'}(x)$ = 0 for $k$ nodes from $x \in $ $\{w_i^j, w_i^{j,1}, ..., w_i^{j, z}\}$.
        \end{itemize}
\end{enumerate} \vspace{3mm}
We compute the value of $M^+(u)$ as follows. \vspace{2mm} \\
$M^+(u) = (\prod_{v \in C(u)} (R(v) \cdot y)) \cdot Q(u)$ \vspace{3mm} \\
Where \vspace{1mm} \\
$y$ = $( M(v) +  A \cdot B \cdot C ) \geq 1?1:0$\vspace{4mm} \\
such that \vspace{3mm} \\
$A = \prod_{w_i \in C(v)} M(w_i)$ \vspace{2mm} \\
$B = \{(\exists_{w_i \in C(v)} \sum_{\substack{w_i^j \in C(w_i) \\ M^-(w_i^j) = 0}} (1-M^{'}(w_i^j))) = k?1:0)\}$  \vspace{2mm} \\
$B$ is set to 1, if and only if, there exists some node $w_i \in C(v)$ with $k$ of its child nodes in $S_v$. \vspace{3mm} \\
$C = ([\sum_{w_i \in C(v)} \{( \sum_{\substack{w_i^j \in C(w_i) \\ M^-(w_i^j) = 0}} (1-M^{'}(w_i^j))) = k?1:0$ + $\exists_{\substack{w_i^j \in C(w_i) \\ M^-(w_i^j) = 0}} ((1-M^{'}(w_i^j)) + \sum_{\substack{w_i^{j,l} \in C(w_i^j) \\ M^-(w_i^{j,l}) = 0}} (1-M^{'}(w_i^{j,l}))) = k ?1:0\} \geq 1?1:0] = |C(v)| ? 1 : 0)$ \vspace{3mm} \\
$C$ is set to 1, if and only if, for each child node $w_i \in C(v)$, either \\(1) $k$ of its child nodes are in $S_v$ or \\(2) there exists some node $w_i^j \in C(w_i)$ that has $k$ of its descendants (including $w_i^j$) in $S_v$. \vspace{3mm} \\
Here, $M^+(u)$ is 1, if and only if, for each child node (say $v$) of $u$, both $R(v)$ and $y$ are 1 and among all the child nodes (say $v$) of $u$, $M^{'}(v)$ is 0 for at most $k-1$ of them. \vspace{5mm} \\
\textbf{Computing $M(u)$:} \vspace{3mm} \\
$M(u)$ = max($M^+(u), M^-(u)$) \vspace{3mm} \\
There exists a \textit{minimum membership dominating set} for $G$, if and only if $M(r)$ = 1. Hence, we obtain the following theorem. \vspace{3mm} \\
\noindent \textbf{Theorem 10.} \textit{The} \MMDS{} \textit{problem on trees is linear-time solvable.} \qed

\section{Conclusion} In this work, we have considered a variant of the dominating set, i.e., the \MMDS{} problem. We have studied the problem on split and bipartite graphs. For the former, we obtained an $\mathcal{O}^*(1.747^n)$ time exact algorithm and for the latter, assuming ETH, we were able to prove the non-existence of a sub-exponential time algorithm for the problem. We also obtain an NP-complete proof for $\Delta = k+2$, when $k \geq 5$, even for bipartite graphs. We have proved that the problem is FPT for the parameter twin cover and for the combined parameter distance to cluster, membership($k$). We have provided a linear-time algorithm for trees. \vspace{3mm} \\
\indent As part of future work, one could look to obtain non-trivial exact algorithms for bipartite graphs (or even for planar bipartite graphs). The lower bound for the problem on planar bipartite graphs (or for planar graphs) is also an interesting open question. The complexity of the problem on graphs with the maximum degree, $\Delta = k+1$ is still open. There is a scope to improve the FPT results obtained for twin cover and the combined parameter distance to cluster, membership($k$). The parameterized complexity of the \MMDS{} problem for distance to cluster and feedback vertex set can be considered. The complexity of the problem for distance-hereditary graphs, cographs and bounded-cliquewidth graphs is open. \vspace{3mm} \\

% \begin{itemize}
% \item Conflict of interest/Competing interests (check journal-specific guidelines for which heading to use)
% \item Ethics approval 
% \item Consent to participate
% \item Consent for publication
% \item Availability of data and materials
% \item Authors' contributions
% \end{itemize}

% \noindent
% If any of the sections are not relevant to your manuscript, please include the heading and write `Not applicable' for that section. 

% %%===================================================%%
% %% For presentation purpose, we have included        %%
% %% \bigskip command. please ignore this.             %%
% %%===================================================%%
% \bigskip

% Springer journals and proceedings: \url{https://www.springer.com/gp/editorial-policies}

% Nature Portfolio journals: \url{https://www.nature.com/nature-research/editorial-policies}

% \textit{Scientific Reports}: \url{https://www.nature.com/srep/journal-policies/editorial-policies}

% BMC journals: \url{https://www.biomedcentral.com/getpublished/editorial-policies}
% \end{flushleft}


\bibliography{sn-bibliography}% common bib file

\end{document}